\documentclass[prd,twocolumn,aps,longbibliography]{revtex4-1} 

\usepackage{url}
\usepackage{hyperref}
\usepackage{mathtools}

\setcitestyle{square}

\parskip=\medskipamount
\begin{document}
\title{Assistive robots for the social management of health: \\A framework for 
robot design and human-robot interaction research}
\author{Meia Chita-Tegmark \& Matthias Scheutz}

\address{Human-Robot Interaction Lab, Tufts University,
200 Boston Avenue, Medford, MA 01255, {\it mihaela.chita\_tegmark@tufts.edu}}

\date{March 2 2020, published 
	in {\it Int.~Journal of Social Robotics},
	\url{https://doi.org/10.1007/s12369-020-00634-z}}

\begin{abstract}
There is a close connection between health and the quality of one’s social 
life. Strong social bonds are essential for health and wellbeing, but often 
health conditions can detrimentally affect a person’s ability to interact with 
others. This can become a vicious cycle resulting in further decline in health. 
For this reason, the social management of health is an important aspect of 
healthcare. We propose that socially assistive robots (SARs) could help people 
with health conditions maintain positive social lives by supporting them in 
social interactions. This paper makes three contributions, as detailed below. 
We develop a  
framework of social mediation functions that robots could perform, motivated by 
the special social needs that people with health conditions have. In this 
framework we identify five types of functions that SARs could perform: a) 
changing how the person is perceived, b) enhancing the social behavior of the 
person, c) modifying the social behavior of others, d) providing structure for 
interactions, and e) changing how the person feels. We thematically 
organize and review the existing literature on robots supporting human-human 
interactions, in both clinical and non-clinical settings, and explain how the 
findings and design ideas from these studies can be applied to the functions 
identified in the framework. Finally, we point out 
and discuss challenges in 
designing SARs for supporting social interactions, and highlight opportunities 
for future robot design and HRI research on the mediator role of robots.
\end{abstract}
%\date{\today}
%\date{January 15, 2020}
\vspace{10mm}	

\maketitle

\section{Introduction}
\label{intro}
Social life is essential for good health 
\cite{cohen2004social,umberson2010social} but often poor health 
detrimentally affects a person’s ability to form and maintain supportive social 
bonds \cite{dejong2006loneliness} leading to a vicious cycle in which health 
and 
well-being are 
impacted negatively. This is especially true for individuals dealing with 
health conditions that require long-term assistance. Whether the impairment 
that restricts social life is physical as in the case of people with 
neuromotor disabilities, cognitive as in the case of dementia, emotional as 
seen in depression, or due to a neurodevelopmental disorder as in the case of 
autism, the effects of an impoverished social life on health range from 
reduced quality of life to reduced life-span \cite{cacioppo2014social}. 

As robots are becoming more common in healthcare, the social management of 
health is an aspect in which their assistance could be extremely valuable. 
Tickle-Degnen et al. \cite{tickle2014collaborative} define the social 
self-management of health as ``the self-care practices that ensure social 
comfort while supporting mental and physical well-being, such as by 
participating in valued social activities, maintaining rewarding interpersonal 
relationships, and seeking help from capable people'' (p.1). 
Socially assistive 
robots (SARs) are machines that are meant to assist users through social rather 
than physical interactions \cite{okamura2010medical}. Developed 
at 
the intersection of assistive robotics and social robotics, the focus of SARs 
is to provide necessary aid for humans and to do so by engaging humans socially 
\cite{rabbitt2015integrating}. In healthcare, SARs are envisioned to 
play roles such as taking medical interviews \cite{van2019social}, monitoring 
and keeping a record of symptoms \cite{briggs2015robots}, 
helping with pill sorting and medication schedules \cite{wilson2016designing}, 
guiding people 
through therapeutic tasks \cite{kim2013structural}, providing companionship 
\cite{banks2008animal}, acting as 
stress reducers and mood enhancers \cite{takayanagi2014comparison}, and 
supporting social interactions 
between humans \cite{wada2007social,arkin2014preserving}. 

In this paper we focus on the last role, that of robots assisting social 
interactions between people. More specifically, we are interested in the 
application of SARs to the social management of health of people with health 
conditions that restrict or negatively impact their social life. We see these 
robots as assistants in breaking the above-mentioned vicious cycle in which 
poor health negatively impacts social bonds, the weakening of which, in turn, 
leads to further decline in health.

Several participatory science studies have shown that people with health 
conditions as well as their caregivers and therapists welcome support from 
robots not just for physical tasks, but also for social interaction. 
For 
example, Williams et al. \cite{williams2019aida} explored ways in which robots 
could augment workers with 
intellectual and developmental disabilities. They observed a group of workers 
with disabilities in the workplace as they 
performed their tasks, and then interviewed some of them about their work 
experience. The study found that among the three most desired features for a 
SAR (as expressed by the workers) was the robot's ability to help 
facilitate more human connection between the workers during work, breaks 
and outside of work.

Another study, by Moharana et al. \cite{moharana2019robots}, focused on 
informal 
careregivers of people with dementia (usually spouses and close family members) 
and their requests in terms of robotic help with caregiving tasks. In addition 
to functions such 
as regulating food intake, prompting and delivering medication, coaching the 
person with dementia through physical therapy exercises and motivating the 
person to be active, caregivers expressed desire for the robot to also support 
interactions between them and the person they were caring for. Caregivers 
wanted robots that 
could facilitate positive interactions with the person they were providing care 
for, such as playing 
favorite songs and inviting both of them to share a dance. They also wanted the 
robot to lessen the emotional stress of the interaction when the person 
requiring care was 
agitated and asked repetitive questions. In this situation, caregivers wanted 
the robot to answer in their place, distract the agitated person, and redirect 
the 
conversation to more enjoyable topics. Finally, since their emotional 
attachment to the person they were caring for made it difficult to deprive them 
of personal 
freedoms, caregivers wanted robots to act as neutral third parties in 
interactions and make the person cared for do things they did not want 
to do, for example 
take their medication, exercise, or stop eating unhealthy things. 

Robot assistance in social interactions is also desired for children with 
disabilities. Most social interactions that children engage in happen in the 
context of play. Introducing structure to play scenarios through robotic 
facilitation can therefore be helpful for children with special needs. Robins, 
Otero, Ferrari, and Dautenhahnm \cite{robins2007eliciting} interviewed a panel of experts comprised 
of therapists, teachers and parents of children with autism to investigate how 
robotic toys can assist social interactions and help children discover 
different play styles, including cooperative play. A recurring theme in the 
panel’s conversation was the need for motivating children with autism to play 
with others, sustain their interest in collaborative play and offer them 
support for how to engage others. Using data from this panel as well as from a 
review of the literature, Robins, Ferrari and Dautenhahn 
\cite{robins2008developing} then explored 
designing robots that could facilitate 
different types of play with therapeutic benefits for children with autism. The 
goal of the project was to design robots that empower children with special 
needs, to prevent isolation and build different skills including social ones.

These findings suggest a few ways in which robots could assist social 
interactions between people for a better social management of health. While the 
other roles for SARs such as providing companionship or coaching focus on 
human-robot interaction, assisting with social life focuses on human-human 
interactions and how robots can provide assistance during the interaction. The 
functions that the robot has to fulfill and the capabilities it needs to have 
to provide effective social support for human-human interactions can be quite 
different from what is required of a robot for successful human-robot 
interaction alone. At this point there doesn’t seem to be a concerted effort 
towards designing robots that can effectively support social interactions 
between people, but such an effort would be highly beneficial for the 
development of SARs that could contribute to the social management of health.

Most of the studies in social HRI focus on the role of the robot as  
interactant rather than as assistant to human-human social interactions. 
However, the field has begun to pay more attention to robots being part of and 
even intervening in social interactions between humans in roles such as group 
member \cite{short2017robot,matsuyama2015four}, facilitator 
\cite{chandra2015can,shen2018stop}, or moderator 
\cite{mutlu2009footing,short2016modeling}. HRI 
studies of robots intervening in human-human interactions vary widely in their 
scope, and are scattered across domains of application, using very different 
robot designs in a variety of context. Some are simply case studies (e.g., 
\cite{giannopulu2012child}), others engage larger participant samples (e.g., 
\cite{wood2015paro}), some studies 
investigate the 
effects of the robot in the context of specific tasks (e.g., 
\cite{short2017robot}), some leave 
the interaction free and open to what participants want to make of it, 
constrained just by the 
robot’s capabilities (e.g., \cite{kidd2006sociable}). Some of the robots used 
are designed with 
clinical applications in mind, such as assisting children with autism (e.g., 
\cite{robins2009isolation}) or providing couple’s therapy (e.g., 
\cite{utami2019collaborative}), but many of them are 
intended for general use, for purposes such as promoting inter-generational 
interactions (e.g., \cite{short2017understanding}). Finally, some of these 
studies were conducted in lab 
settings (e.g., \cite{wood2015paro}) while others in more naturalistic settings 
such as nursing 
homes (e.g., \cite{wada2006robot}). In this paper we draw on this growing, 
although disparate, 
literature (for a summary, see Figure 2), for insights into how robots could 
assist individuals with health conditions in the management of their social 
lives. 

The contribution of this paper is threefold: a) we offer a framework for 
functions that a mediator robot could perform that are motivated by the special 
social needs that people with health conditions have; b) we thematically 
organize and review the existing literature on robots supporting human-human 
interactions in both clinical and non-clinical settings and explain how the 
findings and ideas in these studies fit in the proposed framework; and c) we 
identify and discuss the challenges of designing SARs for supporting social 
interactions between humans. Our framework and the summaries of 
the reviewed studies highlight opportunities for robot design as well as future 
HRI research.

\section{Functions of mediator robots for the social management of health}
\label{framework}
The social lives of people with serious health conditions are different from 
the norm in several important ways. First, people with health conditions can 
have disability-specific difficulties in interacting with others. For example, 
people with Parkinson’s Disease, a neuromotor disorder, might have difficulty 
in expressing emotions in conversations with others due to poor control of 
their facial muscles \cite{tickle2011culture,hemmesch2009influence}, while 
children with autism might have difficulty 
decoding the emotions of others in interactions \cite{harms2010facial}. Second, 
people with 
serious health conditions tend to be more dependent on others for daily 
functioning than their healthy peers and this can shape the types of 
interactions they have within a relationship. For example, people with severe 
health conditions, such as Alzheimer’s disease, in later stages, might need 
round-the-clock supervision and the extent to which they can make autonomous 
decisions about their lives and interactions with others can be limited 
\cite{karlawish2002relationship}. 
Finally, there are types of social relationships that are unique to people with 
chronic health conditions, namely the relationships they form with healthcare 
professionals such as doctors and therapists, and their relationships with 
caregivers. These can pose specific challenges such as forming and sustaining 
fruitful therapeutic relationships \cite{sabat2005capacity}, and adjusting to 
the dynamics of 
caregiver – care recipient relationships, which can often be fraught with 
frustration 
on both sides.

\begin{figure*}
	\centerline{\includegraphics[width=0.75\textwidth]{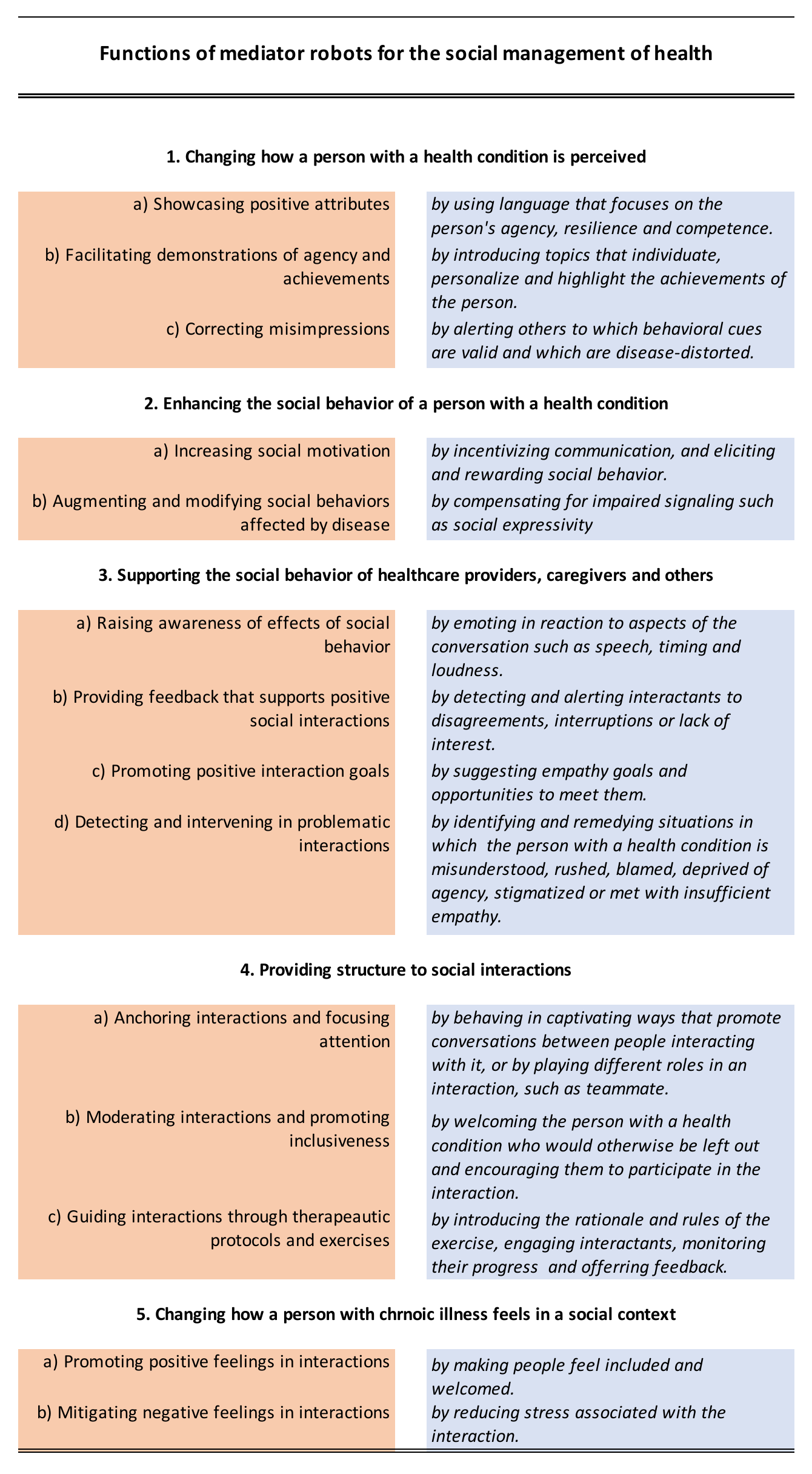}}
	\caption{Summary of the proposed framework with examples of applications 
		(right 
		column).}
	\label{fig:1}       % Give a unique label
\end{figure*}

Given these special social circumstances of people with health conditions, we 
propose that SARs supporting human-human interactions can assist people with 
health conditions in their management of social life by fulfilling these 
functions (for a summary see Figure \ref{fig:1}):

1.	Changing how the person with a health condition is perceived by others 
(e.g., by correcting other’s misconceptions about impairments);

2.	Enhancing the social behavior of the person with a health condition (e.g., 
by supplementing social behavior that the person is not able to convey);

3.	Modifying the social behavior of others towards the person with a health 
condition (e.g., by modeling good behavior or by raising awareness of 
problematic behavior);

4.	Providing structure for interactions between people with health conditions 
and others (e.g., by guiding conversation partners through a therapeutic 
conversation protocol);

5.	Changing how the person with a health condition feels in a social context 
(e.g., by making the person feel listened to or at ease in a stressful social 
interaction).

In what follows we will look closely at each of these functions and explain why 
they are necessary or desirable and how studies in HRI have begun to research 
these functions in robots. We also offer ideas about possible robot design 
directions and gaps in our HRI knowledge.

\subsection{Changing how a person with a health condition is perceived}
\label{sec:2.1}
People react in different ways to a health condition, from impressive 
resilience to major distress, which can profoundly influence the prosocial 
responses they receive from others \cite{preston2013ethology}. 
The way in which people with health conditions are perceived by others can have 
a major impact on their health. In the context of healthcare, how positive an 
impression a patient can make can directly affect how much care they receive. 
Studies have shown that doctors are more inclined to prescribe more care for 
more likable patients. However, doctors seem to be influenced by a patient’s 
perceived traits at an unconscious level. For example, in a study of doctors 
making decisions about Intensive Care Unit (ICU) admissions, the doctors ranked 
the patient’s 
“emotional state” as an important consideration only 6 percent of the times. 
However, when a vignette described a hypothetical patient as being “upbead and 
courageous” as opposed to “sad and discouraged” the same doctors were three 
times more likely to recommend admission to ICU \cite{escher2004national}. Other studies have 
similarly shown that “likable and competent” simulated patients elicited from 
doctors more recommendations for follow-up visits as well as more staff time 
spent on the patient’s education \cite{gerbert1984perceived}. Doctors are not the only ones 
influenced by patients’ character attributes and affect. In a study of 
empathetic responses to naturally-varying affect in real hospital patients, 
participants (who were not medical professionals) watched video-interviews of 
chronically or terminally ill patients talking about their quality of life. 
Participants showed willingness to aid those patients displaying negative 
affect slightly more than those displaying positive affect, but patients 
showing 
little affect were offered the least amount of help.

\subsubsection{Showcasing positive attributes}
\label{sec:2.1.1}
Although there is much opportunity for exploring ways in which robots could 
accentuate one’s positive and empathy-inviting features and behaviors, to our 
knowledge only 
one HRI study has investigated how a robot can change people’s perceptions of a 
person with a health condition. Chita-Tegmark, Akerman, and Scheutz 
\cite{chita2019effects} 
conducted a vignette study in which robots partook in a conversation 
between a patient and a health-care provider: the robot gave a summary of the 
patient’s treatment progress. In doing so, the robot used either  
task-centered language, emphasizing the patient’s level of compliance to the 
treatment plan, or patient-centered language emphasizing the patient’s choices 
and difficulties with regards to the treatment plan. Through its use of 
language, the robot was able to manipulate participants’ impressions of the 
patient: in the patient-centered condition people perceived the patient more 
positively: they thought the patient was more competent, honest and  
self-disciplined rather than disruptive, hostile and disorganized. The same 
results were replicated in other contexts: dieting, learning how to dance or 
job 
training.
Given how important it is for people with health conditions to be perceived in 
a favorable way by others, there is a great opportunity for SARs to positively 
impact these people’s health through social support. SARs could contribute to 
interactions between people with health conditions and others in such a way 
that highlights the positive attributes of the person with the health 
condition. SARs could do this very subtly through choosing language that 
focuses on the person’s agency, resilience, competence etc., like the study 
above has done. 

\subsubsection{Facilitating demonstrations of agency and achievements}
Another way for robots to influence how a person with a health condition is 
perceived is to introduce in conversations topics that 
individuate, personalize, and highlight the achievements of the person. To 
humanize patients, Haque and Waytz recommend that, at a minimum, reminders be 
offered to the medical professionals and others about the patient's past or 
present profession, hobbies and family 
life \cite{haque2012dehumanization}. Additionally, creating opportunities to 
reflect on the creative overcoming of 
challenges caused by the health condition, instead of the impairments 
associated 
with it, can be a fruitful way of changing for the better the way the person 
with the health condition is perceived. This is especially important for 
interactions between patients and healthcare providers, which tend to be 
focused on the disease and its negative effects on the patient, with little 
room for discussing the patient’s achievements and thus with little opportunity 
to observe the patient exhibit positive affect.

\subsubsection{Correcting misimpressions}
Additionally, it is often the health condition itself that leads to negative 
impression formation. For example, people with Parkinson’s Disease are often 
perceived to be less extraverted and more neurotic 
\cite{tickle2004practitioners} and, if a woman, as less supportive  
\cite{hemmesch2009influence}. This is due to a symptom of Parkinson’s Disease 
called 
facial masking, 
which affects facial muscles and facial expression. In these situations, in 
which the health condition is the root cause of the misimpression, SARs could 
intervene by correcting misconceptions and alerting people to which behavioral 
cues are valid, 
and which are not. In the context of Parkinson’s Disease, 
for example, SARs could instruct interactants to pay attention to what the 
person with Parkinson’s Disease is saying as a better indicator of their 
personality and mood, rather than their facial expression, which is affected by 
the disease \cite{lyons2004impressions}. In addition to 
supporting others in forming better impressions of people with health 
conditions, SARs could also 
assist people with health conditions by compensating for a variety of social 
impairments caused by the health condition itself.

\subsection{Enhancing the social behavior of a person with a health condition}

\label{sec:2.2}

Many health conditions can affect a person’s ability to engage in positive 
social behaviors. A disorder that has received much attention from the robotics 
community is Autism Spectrum Disorder (ASD). Social impairments are a core 
symtpom of ASD, a neurodevelopmental disorder affecting 
1 in 59 individuals \cite{baio2014prevalence}. ASD is characterized by 
persistent social 
deficits across multiple contexts, such as: abnormal social approach, failure 
to 
initiate and respond to social interactions, abnormalities in in eye contact 
and body language, difficulties in sharing imaginative play or absence of 
interest in peers. Several case-studies have documented the potential for 
robots to support social behavior in children with ASD by incentivizing 
communication and evoking, eliciting rewarding and reinforcing social behavior.

\subsubsection{Increasing social motivation}
\label{sec:2.2.1}

Giannopulu and Pradel \cite{giannopulu2012child} have documented a case of a 
child with 
autism using a robot as a mediator for his interaction with a therapist in a 
free play scenario. The robot had a very simple design: a schematic face-like 
cover made of geometric shapes (circles for eyes and mouth, and triangle for 
nose) on top of a remote-controlled locomotion hardware, able to move forward, 
move back and swivel. An operator manipulated the robot wirelessly in the 
following way: if the child approached, the robot moved back; if the child 
moved 
away, the robot followed the child; and if the child was motionless, the robot 
turned itself around to grab the child's attention. After establishing an 
interaction with the robot, the child began to use the robot to express 
positive emotion, an interaction cue directed at the therapist. When the child 
interacted in a standalone 
manner with the robot, the positive emotion expression was quasi-absent, 
leading the authors to believe that the expression of enjoyment was the 
indication of a `passage' from child-robot interaction to a child-therapist 
interaction. The authors interpret this as an indication that the child was 
using the robot as a tool for human-human interaction and that the interest 
elicited by the robot was an essential stepping stone for facilitating the 
interaction with another person.

Robins, Dautenhahn and Dickerson \cite{robins2009isolation} described three case studies conducted 
with minimally verbal, low functioning children with autism. In the studies, a 
humanoid robot facilitated interactions between these low functioning autistic 
children and other people. Notable behaviors that the children engaged in 
included reaching for the experimenter’s hand, which was surprising to both the 
experimenter, parent and therapist given the autism severity of the child. 
Another example of engaging in social behavior in the context of playing with 
the robot was exploring the teacher’s eyes and face after exploring the robot’s 
eyes and face as well as sharing excitement with the teacher by reaching out to 
her and asking her to join in the game. Finally, a child was gradually able to 
participate in an imitation game with the therapist taking turns controlling 
the robot and imitating the robot. Through this game the child learned to look 
at the therapist to see how she imitated the robot. Eventually the child was 
able to successfully engage in the same imitation game with another child. The 
authors argue that the robot allowed the children to demonstrate some 
interactional competencies and generalize this behavior to the co-present 
others.

Beyond case studies, Kim et al. \cite{kim2013social} showed that in a structured play 
interaction, children with autism spoke more with an adult confederate when the 
interaction partner was a robot than when it was another human or a computer 
game. The researchers used Pleo, a dinosaur shaped robot which was programmed 
to show interest in different objects and exhibit positive and negative 
emotions. The children were excited and interested in the robot and were thus 
motivated to ask how the robot works, whether it “was real” and what the robot 
was doing. The authors suggest that the inclusion of the robot in the task 
can 
thus serve as an embedded reinforcer of social behavior.

In the case of autism elicitation and maintenance of social behavior is a 
challenge specific to the disorder and robots can help by increasing social 
motivation and evoking and reinforcing social engagement. These robot functions 
are also generalizable to other health conditions. For example, this type of 
assistance might also be useful for people with depression or anxiety where 
social behavior might be absent or insufficient because of emotional 
difficulties \cite{segrin2000social}.  

\subsubsection{Augmenting and modifying social behaviors affected by disease}
\label{sec:2.2.2}

In the context of other health conditions robots might be useful in enhancing 
social behavior not by eliciting more of it, but by modifying or adding to it 
in specific ways. For example, in the case of Parkinson’s Disease, it has been 
proposed that a robot could be used to convey emotions that the person with 
Parkinson’t Disease is incapable of expressing due to facial masking \cite{tickle2014collaborative}. Arkin and Pettinati \cite{arkin2014moral} have proposed the development 
of a robot co-mediator that would increase the emotional communicative 
bandwidth of the person with PD in such a way that would facilitate empathic 
response in a caregiver. The robot would express through body motions and 
postures the mental states of the person with Parkinson’s Disease with the goal 
of eliciting empathy when incongruences arise between the mental state of the 
person with Parkinson’s Disease and the other interactant.

Most of the studies on how robots can help enhance the social behavior of 
people with health conditions are observational case studies or conceptual 
proposals. More HRI studies are needed to determine how robots can address 
social interaction needs that are specific to various health conditions. Most 
of the studies in which robots help with social interactions focus on 
autism, but there are many other health conditions that negatively impact the 
ability to engage in effective and appropriate social behavior that SARs could 
assist with. However, in social interactions it is not only the social behavior 
of the person with the health condition that matters, but also that of the 
interaction partner. Robots could provide support for those interacting with 
people with health care conditions with the aim of making such relationships 
stronger and more positive.

\subsection{Supporting the social behavior of healthcare providers, caregivers 
and others}
\label{sec:2.3}

In social interactions people with health conditions run the risk of being 
reduced to their impairments. In relationships with others, especially with 
those that provide care, they can be seen almost exclusively through the lens 
of their needs, which can harbor dehumanization. Specifically, people with 
health conditions may be treated less like persons and more like objects or 
nonhuman animals \cite{haque2012dehumanization,haslam2014dehumanization}. It 
is not 
that empathetic and humanizing care is not an aspiration of those providing it; 
in fact, it very much is, but often dehumanization ensues because of the need 
of 
health care providers and caregivers to create distance and emotional barriers 
to protect themselves from the emotional drain ensued by dealing with health 
care problems on a daily basis 
\cite{haque2012dehumanization,decety2010physicians,cheng2007expertise}. 
Caregiving relationships can be 
emotionally taxing and accompanied by frustration, thus in spite of best 
intentions, the social behavior of those providing care can often lack in 
empathy. 
However, empathy and humanization of care has been shown to be beneficial for 
health outcomes and many studies highlight the importance of empathy and 
patient-centered approaches in medical practice 
\cite{rathert2013patient,stewart2001towards,constand2014scoping}. It has been 
proposed 
that admissions for medical school be based on empathy and emotional 
intelligence aptitudes \cite{haslam2007humanising}, and that training in 
empathetic behavior be 
required for health care professionals \cite{riess2014empathy}. 

SARs could be used to support health care providers and caregivers when 
interacting with people with health conditions to ensure that dehumanization is 
avoided. Based on studies in HRI so far, we propose four main ways in which 
SARs could support the social behavior of health care providers and caregivers: 
a) by raising awareness of one’s social behavior and its effects on others, b) 
by providing feedback that supports empathetic behavior, c) by helping people 
set and maintain empathy goals for their interactions, and d) by detecting and 
intervening when problematic interactions occur.

\subsubsection{Raising awareness of effects of social behavior}
\label{sec:2.3.1}

A first requirement for self-correcting one’s problematic social behavior is 
being aware of it and of its effects on others. However, oftentimes people 
remain oblivious to what they are doing and how it affects those around. 
Hoffman et al. \cite{hoffman2015design} used an 
emoting and empathy-evoking robot, Kip1, to increase awareness of the effect of 
one’s behavior in an interaction. The robot monitored nonverbal aspects of the 
conversation (speech, timing, silences and loudness) and responded with a 
gesture indicating curious interest when the conversation was calm and a 
gesture indicating fear when the conversation was aggressive. They used the 
robot as a peripheral companion in conflict conversations between couples. 
Couples were asked to discuss a topic they had high disagreement about in the 
presence of the robot. After the interaction, couples reported the same level 
of comfort in conversing next to the reactive robot as to the control, 
non-reactive robot which did not behave in response to their conversation. 
Also, couples attributed social human characteristics to the reactive robot. No 
quantitative data was reported on how the robot’s reactions might have changed 
the conversation, but a qualitative account suggests that couples sometimes 
reacted to the robot’s gesturing by adapting their own behavior, for example, 
pausing and taking the conversation in a different direction. Such capabilities 
in robots could also be used in the context of caregiving. This could assist 
health care providers and caregivers in monitoring their own social behavior 
and 
correcting unintended, dehumanizing or unempathetic aspects of the 
interaction. 

\subsubsection{Providing feedback that supports positive social interactions}
\label{sec:2.3.2}

A step further in assisting people with the management of their social behavior 
is to provide feedback that supports positive social behavior. Tahir et al. 
\cite{tahir2014perception} used a Nao robot for providing real-time feedback to 
participants in a 
dyadic conversation. The Nao sensed and recorded conversational cues (e.g., 
number of natural turns, speaking percentage, interruptions etc.) and prosodic 
cues (e.g., amplitude) and then used machine learning algorithms to determine 
the social state of the participants (level of interest, agreement and 
dominance). Based on its model of the participants' state, Nao would alert the 
speakers when their voice was too high or too low or when the conversation was 
problematic due to too many disagreements or interruptions. The robot provided 
sociofeedback, alerts through speech accompanied by body postures in the 
following situations: when the conversation partners seemed uninterested in the 
discussion (``You both seem uninterested.''), when one person was speaking too 
much (``You are talking a lot.''), when one person was being too aggressive 
(``Please calm down.''), when someone's voice was too loud (``Please lower your 
volume.'') or not loud enough (``I am sorry, I cannot hear you.'') and when the 
conversation was proceeding normally (``Good, carry on.''). To validate the use 
of the robot as a social mediator, participants were asked to produce certain 
behaviors such as talk too loud, too much or to interrupt frequently. 
Participants felt that Nao's performance was good in terms of clarity: whom it 
was addressing and what it was saying. In terms of timing, some participants 
felt interrupted by the Nao. Most importantly participants indicated that they 
liked receiving socio-feedback from Nao and voted the Nao as their second 
favorite platform for receiving sociofeedback after virtual humans. 

As opposed to the study by Hoffman et al.  \cite{hoffman2015design}, in which 
the robot had a 
peripheral role in the interaction, in this study the robot intervened in the 
conversation. Also, while in the study by Hoffman et al. the robot’s behavior 
was evocative, in this study it was evaluative. Although the results of the 
study seem promising (participants reported favorable impressions of the robot 
and a desire to receive sociofeedback), it is unclear how welcome the 
sociofeedback would be in a real interaction, one in which behavior was not 
acted, especially when the robot points out undesired behavior. People might 
feel uncomfortable having their interaction evaluated in this manner by the 
robot. 

Although research remains to be done to determine the ecological validity of 
this particular approach, the general idea of having robots infuse interactions 
with supportive social cognitions through sociofeedback merits further 
attention. In the context of caregiving, sociofeedback could help rapidly 
deescalate tense interactions and further encourage positive ones. The nature 
of the sociofeedback could be adjusted to the specific problems encountered by 
the caregiver and the robot could even act as an emotion regulation tool. 
Moharana et al. \cite{moharana2019robots} recounts the desire of a caregiver 
who wanted a robot 
that could remind her that her husband’s anger was not because of her poor care 
towards him but because of his dementia. Such reminders could be incorporated 
in the sociofeedback given during an interaction. Also, the sociofeedback need 
not be primarily negative. Activating positive 
social cognitions could be useful as well, for example the robot 
could point 
out how attentive the conversation partner is, how excited she is about the 
topic, or 
how much joy it brings her to be part of the interaction. Such cognitions 
could perhaps be empathy-inducing for the caregiver and humanize the 
person receiving care.

\subsubsection{Promoting positive interaction goals}
\label{sec:2.3.3}

Another way in which robots could support caregivers is by helping them set and 
maintain positive goals for their interactions. This could be highly beneficial 
in care scenarios especially in interactions that have competing and perhaps 
even conflicting goals, for example, making sure a person with dementia takes 
their medication on time, while also maintaining a patient, tolerant attitude 
in the face 
of their forgetfulness. Wilson, Arnold and Scheutz \cite{wilson2016relational}  
have developed a framework for evaluating the design of human-robot 
relationships when tradoffs appear between the succesful completion of task, 
and 
the maintainance of positive relationships with the human user. This framework 
could be adapted to scenarios involing robot mediation of human-human 
interactions that require the balancing of different types of goals.

Short and Matarić \cite{short2017robot} used robots as mediators in 
collaborative tasks, which influenced the interactions by promoting different 
types of goals. They developed two algorithms to specify the robot's behavior: 
one in which the robot suggests goals that are optimal from a 
performance-maximizing standpoint (performance-reinforcing) and an algorithm in 
which the robot suggests goals that the poorest-performing team member can help 
accomplish (performance-equalizing), thus increasing the collaborative 
contribution of this member. 
Contrary to their hypothesis they found that group cohesion was higher in the 
performance-reinforcing rather than the performance equalizing-condition. 
Group performance was also higher in the performance-reinforcing condition. 
They also found that the more a robot spoke to a participant, the higher the 
group 
cohesion they reported and the more they helped the other participants in the 
group. Participants completed over half of the robot's suggestions, although as 
the authors note there are further opportunities for improving the timing and 
salience of the robot's suggestions. Also, participants took more of the 
robot's advice in the performance-reinforcing condition than in the 
performance-equalizing condition. After the task, participants' attitudes 
towards robots on the \textit{Attitudes towards Situations and Interactions 
with Robots} subscale of the Negative Attitudes towards Rorobts Scale became 
more negative.

The findings of this study are particularly promising because they clearly show 
that robots can modify people’s social behavior in interactions. Additionally, 
the study develops and tests two different ways in which the robot could 
behave. This is important because further development of SARs for the social 
management of health will require a lot of fine-tuning and personalization of 
the robot’s behavior to meet the specific needs of the user, determined by the 
user’s particular health situation as well as personality and preferences. 
Through future research, it will be important to understand which suggestions 
or 
types of suggestions people readily take from robots and which they ignore. 
Also, a cause for slight concern is that participants seemed to have a more 
negative attitude towards the robot after completing the task, thus it will be 
important to understand how that would affect long-term use.

\subsubsection{Detecting and intervening in problematic interactions}
\label{sec:2.3.4}

Finally, SARs could help detect and intervene in problematic interactions 
between people with health conditions and their caregivers or health care 
providers. The idea is that when an interaction becomes problematic and a 
person with a health condition is misunderstood, rushed, blamed, deprived of 
agency, stigmatized, or met with insufficient empathy, the robot would 
intervene 
to remedy the situation. The robot’s intervention could take different forms, 
focusing on adjusting the behavior of the person with the health condition as a 
way of helping the caregiver, focus on adjusting the caregiver’s behavior or 
both.

Shim, Arkin, and Pettinatti \cite{shim2017intervening} implemented and evaluated a mediator robot 
that intervenes in situations that might lead to the stigmatization of people 
with health conditions. Their approach was to focus on modifying the behavior 
of the person with the health condition, however, evaluative data from 
participants indicated that this might not be the preferred approach. The 
researchers implemented an intervening ethical governor model onto a robotic 
platform (the Nao robot), which models the relationship between the patient and 
caregiver, detects discordances between the patient’s level of embarrassment 
and the caregiver’s level of empathy, and intervenes through speech and 
movement to correct these gaps in communication and incompatibilities between 
emotional states. The researchers devised four different scenarios illustrative 
of four ethical rules of interacting: prohibition of angry outbursts from the 
patient, prohibition of withdrawal from the patient, obligation of the patient 
to stay in the therapeutic activity/session, and the obligation of the patient 
to follow safety requirements. Four videos were recorded of acted problematic 
interactions illustrating the intervention of a mediator robot who followed the 
rules above. Qualitative data was obtained from nine elderly participants who 
were shown the videos and who were guided through standardized open-ended 
interviews about the scenarios depicted in the videos. Participants felt that 
the most appropriate and essential type of intervention of the robot was the 
one corresponding to the “safety-first” rule, in which the robot made sure the 
patient follows safety requirements. Participants had a negative reaction to 
the robot’s intervention in the other scenarios, feeling that the robot sounded 
judgmental, commanding and critical of patients, which was deemed unacceptable. 
In the videos, the robot always addressed the patient rather than the caregiver 
and the rules referred to the patient’s behavior rather than that of the 
caregiver. Participants indicated that it would be more appropriate for the 
robot to indicate to the caretiver situations needing intervention. The robot 
should do this in a subtle way and then allow the caregiver to remedy the 
situation instead of the robot intervening.

Further research is clearly needed to establish the best ways in which robots 
could intervene in problematic situations. As we have seen, the robot 
intervention itself can increase the feeling of blame and criticism, which was 
perceived as unacceptable. Also, as participants imply when talking about their 
preference for the caregiver to handle the remediation, some actions might be 
seen as appropriate coming from a human interactant but not from a robot. An 
example, perhaps not of an appropriate intervention per se in the social 
management of health context, but of a study that has systematically attempted 
to compare human with robot intervention is \cite{stoll2018keeping}.
Stoll, Jung and Fusell \cite{stoll2018keeping} studied the use of humor by 
robots for 
conflict mitigation. Humor has been shown to alleviate tension in interpersonal 
conflict, which makes it a commonly used strategy for diffusing conflict \cite{bippus2003humor}. Participants watched videos of robots or humans using humor to 
diffuse a conflict situation between two roommates. Although affiliative and 
aggressive humor was perceived as less appropriate when used by a robot rather 
than a human, self-defeating humor was well received from both. Unfortunately, 
the study does not report how effective people felt the humor was at diffusing 
conflict.

Oftentimes the behavior of both interactants needs to be adjusted for a 
problematic situation to be remedied. A study by Shen, Slovak and Jung \cite{shen2018stop} 
offers an example of how a robot could intervene and guide the remediation of a 
problematic situation. Principles from this study could be extended and adapted 
to applications in the context of caregiver-care recipient relationships. Shen, 
Slovak and Jung used a mediator robot to support children in resolving 
interpersonal conflicts constructively. What is interesting about this robot is 
that its actions were programmed around formalized steps from a conflict 
negotiation procedure: Teaching Students to be Peacemakers (TSP). Examples of 
steps are: stating what you want and giving your underlying reason (``I 
want...because...'') or expressing how you feel (``I feel mad or sad.'').  
The 
robot facilitated the conflict resolution by identifying when a conflict was 
happening, alerting the children and then guiding them through the negotiation 
steps by using prompts matched to the protocol steps, such as: ``Telling each 
other what you want/how you feel can help. Can you try that?''. This robot was 
operated in a Wizard-of-Oz manner, so more development is needed in terms of 
making the robot autonomous and robust to the messiness of natural dialogue. 
Attention should be paid to proper timing and pacing so that the robot can 
intervene at the right time and follow an appropriate progression through the 
protocol steps.
Using protocols for supporting interactions 
can, 
however, be a very fruitful approach for designing mediator robots, because of 
the scripted nature of 
conversation protocols, which are easier to handle by robots. Conversation 
protocols are good tools for structuring interactions. In the following section 
we summarize and expand on studies which have investigated how robots can 
provide structure to interactions through conversation protocols and other 
methods.

\subsection{Providing structure to social interactions}
\label{sec:2.4}

Providing structured interactions for people is perhaps the most valuable way 
in which SARs could support the social management of health. People with health 
conditions, especially the elderly, are at high-risk for isolation, 
which can have serious detrimental effects on health \cite{cacioppo2003social}. 
It is thus 
valuable for SARs to create opportunities for people with health conditions to 
interact with others and participate fully in social life. Structuring social 
interactions in ways that make it easier for people with health conditions to 
join in and follow along is thus crucial. There are different 
levels, of increasing complexity, at which SARs could structure social 
interactions for people: a) by serving as the focus of attention and anchoring 
the interaction, b) by moderating an interaction, providing participation 
opportunities through speech and acts of encouragement, and overall promoting 
inclusiveness, and c) by guiding people through standard interaction protocols 
or exercises.

\subsubsection{Anchoring interactions and focusing attention}
\label{sec:2.4.1}

The lowest level of structure for an interaction is offering anchoring, serving 
as a point of focus and through that creating an opportunity (or an excuse) for 
interaction. To accomplish this, the SAR does not need to have very 
sophisticated capabilities, it simply needs to behave in a way captivating 
enough that it prompts conversation between people interacting with it. This 
low-level support for structuring human-human interactions by robots has 
already been fairly widely explored especially with older adults.

Wada and Shibata \cite{wada2006robot}  used the Paro robot in a care-house for the elderly in 
Japan. Paro is a pet-like robot in the form of a seal pup which responds to 
sounds and touch by making noises and moving. The robot was placed in a public 
area where the residents of the house could meet to interact with each other 
and was activated for 9 hours every day. The researchers found an increase in 
density of the residents’ social networks after the introduction of Paro, which 
suggests that the robot stimulated communication among residents, strengthening 
their social ties. Additional data from this research project presented by Wada 
and Shibata in \cite{wada2007living} showed that the time residents spent in 
the public area 
increased after the introduction of Paro. Qualitative data suggest that 
residents who felt impaired in their communication due to speaking in a 
different dialect found Paro useful in breaking down this communication barrier 
and felt more comfortable talking to others. Additionally, caregivers and 
residents remarked that the topics talked about became more positive when Paro 
provided an anchoring for the conversation.

In the United States, Kidd, Taggart, and Turkle \cite{kidd2006sociable} used the Paro robot in 
two nursing homes to investigate whether robot interactions generated more 
social activity. People who interacted with Paro in its “On” mode had more 
social interactions and this effect was further increased by the presence of 
caregivers or experimenters participating in the interactions. The authors 
conclude, drawing also from previous experience with using robots in nursing 
homes, that robots could be useful at stimulating small group engagement and 
could be a beneficial addition to the very impoverished social setting of 
eldercare facilities, which usually consists of the TV room where people, even 
if in each other’s presence, do not engage in conversation with each other. 

Robinson, MacDonald, Kerse and Broadbent \cite{robinson2013psychosocial} also 
used the Paro robot in a 
residential care facility and compared its effect on social interactions with 
the effect of an actual pet. The facility benefited from visits from a dog 
belonging to the activities coordinator. The behavior of the residents was 
observed during various activities, during the dog’s visit and during group 
interactions with the Paro robot. Observations showed that more residents were 
involved in discussions about the robot in comparison to discussions about the 
resident dog, and the robot appeared in more conversations amongst residents 
and with staff members than the dog. This could simply be due to the fact that 
no special activities were organized around the dog, while group gatherings to 
interact with Paro were organized, even though the specific way in which 
participants interacted with the robot was not prescribed.

For a more systematic (although perhaps less ecologically valid) investigation 
of Paro’s effects on social interactions, Wood, Sharkey, Mountain and Millings 
\cite{wood2015paro} conducted an in-lab study using the Paro robot for social mediation in 
human-human interactions. Participants were asked to interact with the robot 
together in any way they wanted to. The study presents more direct, 
quantitative data on the effects of the robot on social interactions. 
Participants in the active Paro condition (the robot being "On") rated the 
quality of the interaction 
and the enjoyment of interacting with the other person as higher. Although Paro 
is not designed specifically to encourage interaction between people, the 
robot’s social mediation effect likely came from serving as a focus for the 
interaction.

Paro, is not the only robot that has been used to elicit human-human 
interactions. Joshi and Šabanović \cite{joshi2019robots} investigate the use of a variety of 
robots for stimulating intergenerational interactions in a nonfamilial setting: 
a co-located preschool and assisted living center for older individuals with 
dementia. They used four commercially available robots: Paro, Joy for All, Nao 
and Cozmo, which have different capabilities. Paro and Joy for All are pet-like 
robots that react to being held or stroked. Nao is a humanoid robot that can 
speak, move and track people, and Cozmo is a palm-held robot that can drive, 
speak in short sentences and express emotions. The experimenters worked in 
collaboration with the preschool and assistive living center staff to design 
activities that would lead to interactions between the residents and the 
preschoolers, customizing for the values and goals promoted by the center: 
increased inter-generational contact,  increased peer engagement, meaningful 
interactions for both adults and children, opportunities to collaborate and 
share, and reduced need for outside management of the activity. By observing 
the behavior of the participants during the interactions, the experimenters 
found that activities involving robots were often able to provide more 
opportunities for intergenerational interactions than other types of activities 
such as drawing, puzzle solving and making music, and also required less 
intervention from staff members. The best robots for inter-generational 
interactions were Paro and Joy given their slow pace for responding which 
prevented older adults from getting overwhelmed and made the children impatient 
and inquisitive, giving the older adults opportunities to interact with the 
children. The Cozmo robot, although it facilitated peer interactions among 
children was not engaging for the older adults. The study is a great example of 
possibilities for introducing robots that can enhance interactions in 
real-world settings by working closely with the community members involved.

Robots’ abilities to stimulate social interactions has also been studied with 
children with autism. Werry, Dautenhahn, Ogden, and Harwin \cite{werry2001can} used a mobile 
robot in dyadic play interactions between children with autism. They observed 
three pairs of children interact with the robot and with each other, and 
concluded that by serving as a focus of attention, the robot facilitated 
interesting types of interaction structures between children, such as 
instruction, cooperation and even possibly imitation. This was one of the first 
observational studies exploring interaction structures in autism afforded by 
the introduction of robots as an anchor for human-human interactions. 

A more sophisticated way of anchoring and eliciting interaction between people 
is to go beyond using the robot simply as an attention focus, and instead have 
a robot play different active roles in an interaction. Given the current 
limitations of robots, and the fairly narrow number of tasks any given robot 
can perform, games can be a suitable context in which mediator robots can be 
used. Short et al. \cite{short2017understanding} studied family groups as they played games with a 
robot, with the goal of improving intergenerational family 
interactions. The robot played different roles depending on the game, being a 
competitor, a performer (one game consisted of working as a team to make the 
robot dance), or supporter - making positive comments about the family's 
collective creation in a scrapbooking creative game. Unfortunately, the study 
does not explicitly measure how specific robot behaviors affected the 
interaction between family members. The study was instead focused more on how 
the 
different group members perceived and interacted with the robot and their 
engagement with and thoughts about the games. However, this study is a great 
example of a protocol that could be used to study robot support for “gamified” 
interactions. For people with health conditions, especially for children with 
health conditions, therapeutic game-play supported by SARs can be a motivating 
way to develop and practice social skills. 

\subsubsection{Moderating interactions and promoting inclusiveness}
\label{sec:2.4.2}

The studies explored so far in this section focus on increasing the motivation 
of people to participate in social interactions. However, even when the 
motivation to interact exists, people with health conditions often encounter 
challenges in terms of entering ongoing interactions and keeping up with them. 
For example, people with Parkinson’s Disease, due to slowness of speech and 
word-finding difficulties, have a hard time entering a conversation or keeping 
up with the rapid pace of one \cite{miller2006life,mcnamara2003pragmatic} . 
Children with autism have difficulties 
producing appropriate social behaviors to initiate and maintain social 
interactions \cite{rogers2000interventions}. People with social anxiety or 
simply people that are 
unusually shy can also have a difficult time to get a piece in edgewise in a 
conversation. SARs could support these people by moderating social 
interactions, offering assistance for conversation and group entry, and 
generally promoting social behaviors that lead to inclusiveness.

For example, Short, Sittig-Boyd and Matarić \cite{short2016modeling} used a robot to moderate a 
group storytelling activity. The robot kept track of participation (how much 
each group member spoke) and asked general or specific questions at fixed time 
intervals to the participant with the least speech in the last time interval. 
Each group participated in the task twice, one time with the robot as moderator 
and one time with the robot as ``active listener'' - the robot watched the 
speaker and produced an utterance such as ``huh'' or ``okay''. They found 
marginally significant results for an increase in group cohesion in the 
moderated condition and increased speech in the moderated as opposed to the 
unmoderated condition.

Another example of study in which a robot was used to promote conversation 
inclusiveness was conducted by Tennent, Shen and Jung  \cite{tennent2019micbot} who used a 
peripheral robotic object to increase group engagement and also to improve 
problem solving performance. They designed a robotic microphone that exhibited 
two engaging behaviors: following – turning towards the person speaking, and 
encouraging – rotating towards the participant who spoke the least and leaning 
towards that participant as an invitation to speak. The authors found that the 
robotic device, when operating according to the above described engagement 
algorithm, increased evenness in backchanneling: namely the participants took a 
more even number of turns to engage in active listening of one-another. The 
evenness of group backchanneling turns then significantly predicted 
problem-solving performance on the Desert Survival task (participants were 
discussing the rank order of 15 most useful items for surviving in the desert, 
their response as a team being compared to that of experts).

These studies show that speech, and even minimal non-verbal gestures can be 
successfully used by robots to promote inclusion of others in social 
activities. Furthermore, Mutlu et al. \cite{mutlu2009footing} have shown 
that robots with fairly low capabilities can be effective in shaping the roles 
of people in conversations: as addressees, bystanders or overhearers. Through 
gaze cues alone, by looking or not looking at the participant when talking, the 
robot was able to manipulate who participated and attended to a conversation as 
well as the participant’s feelings of groupness and their liking of the robot. 
Participants to whom the robot communicated the role of addressee attended to 
the task more and felt stronger feelings of groupness. Participants whose 
presence was acknowledged by the robot, those in the role of addressee or 
bystander liked the robot more.

A more detailed investigation into the specifics of how a robot should act to 
make sure people can participate meaningfully and equally in conversation is 
described by Matsuyama et al. \cite{matsuyama2015four}. They used a robot 
for facilitating a conversation between three participants in which two 
participants had a strong engagement with each other evidenced by lots of 
back-and-forth conversation turns, and one of the participants was left out 
(side-participant). The robot acted as a fourth participant to the conversation 
and its goal was to ``harmonize'' the conversation, by engaging the person left 
out. The robot had to detect the strength of the engagement between 
participants and identify the participant who had a side role. Then the robot 
intervened to include the unengaged participant. Videos were recorded of 
conversation scenarios and participants were asked to rate the appropriateness 
of the robot's behavior, the feeling of groupness and the timing of the robot's 
intervention. The robot intervened in the conversation either by directly 
addressing the participant who was left-out or by initiating a procedure: first 
addressing a comment to one of the engaged participants (i.e., claiming an 
initiative), waiting for a response (i.e., approval of the initiative) and then 
yielding the floor to the left-out participant. In this process, the robot 
either maintained the topic of conversation or initiated a new topic. 
Participants felt that the robot behaved most appropriately and there was a 
stronger sense of groupness when the robot attempted to include the 
side-participant by initiating a procedure without shifting the topic of 
conversation. Participants felt that intervening after two rounds of 
back-and-forth between the engaged participants was more appropriate than after 
the first round.

These studies demonstrate that robots can meaningfully moderate interactions to 
encourage the inclusions of people who would otherwise be left out. All these 
studies were conducted with healthy participants, but the robot design 
features presented can be applied also to the social management of health, 
addressing the needs of people with health conditions for participating more 
fully in social life. Further research is needed to determine what adjustments 
in the robot behavior might be needed to address specific needs related to 
health conditions. For example, robots might need to engage in additional 
special behavior in order to slow down a conversation to make sure someone with 
poor processing capacities has enough time for comprehension.

\subsubsection{Guiding interactions through therapeutic protocols and exercises}
\label{sec:2.4.3}

The highest level of interaction structuring that SARs could provide is to 
guide people through structured interaction tasks or protocols. Therapeutic 
programs often incorporate structured interaction exercises, which are easier 
for robots to handle than free dialogue. SARs could be used as facilitators of 
such therapeutic exercises focused on improving interactions between people as 
a supplement and reinforcer to human-delivered therapy. For example, Utami and 
Bickmore \cite{utami2019collaborative} explored robot-driven couples counseling using a humanoid 
robotic head. The robot was operated in a Wizard-of-Oz manner and it guided 
couples through a rapport-building task and two counseling exercises: a 
gratitude exercise in which the couples were asked to recall and share three 
recent positive behaviors of their partner and the Caring Days exercises 
(commonly used in the Behavioral Couples Therapy) in which each partner made a 
request for a behavior that the other member of the couple could perform to 
show that they cared. The robot explained the rationale for the exercises, 
asked the couples to engage in the exercise and provided feedback. The study 
found a significant decrease in participant’s negative affect post-intervention 
and a significant increase in self-reported intimacy. The couples indicated 
that they enjoyed the interaction with the robot and with each other and they 
rated their partner’s responsiveness as high. Also, intimate behaviors such as 
touching and comforting were observed during the session. The post-session 
open-ended interviews revealed interesting insights about people’s experience 
with 
the robot. Participants felt that the robot’s responses were very generic and 
that the interaction was too structured, which could perhaps be improved in 
future iterations of the study by having the robot engage in some naturalistic, 
random behavior extraneous to the task. However, what is encouraging is that 
even though participants thought that a human counselor would be more genuine 
and better at understanding non-verbal behaviors (such as facial expressions) 
some participants felt that the advantage of the robot was its ability to stay 
non-judgmental and unbiased. Also, very promising is that participants 
indicated that the interaction with the robot was preferable to reading 
self-help material and practicing exercises by themselves. They recognized the 
robot as being helpful in structuring the interaction as a “neutral third 
party”. Even couples who were familiar with the skills practiced with the 
robot liked being reminded of them. Using SARs for therapeutic exercises 
like these which could also be relevant for 
strengthening the bonds between caregivers and care recipients are 
very much in line 
with what participants in the study by Moharana et al. \cite{moharana2019robots} expressed: a 
desire for robots to help accentuate positive shared moments with the person 
they were caring for and act as neutral parties to diffuse tension when 
unwanted tasks 
needed to be completed (e.g., adherence to treatment). Guidance through 
structured interactions can be used not just for creating positive connections 
but also for remedying strained ones. We have already discussed in the previous 
section the study by Shen, Slovak and Jung \cite{shen2018stop} which is an example of an 
interaction protocol for conflict resolution. 

Finally, robots can assist people assist others by guiding them through 
assistance-giving protocols. Many caregivers are family members, not trained 
professionals, and it can often be difficult for non-professionals to gauge
the right amount of support needed by the person requiring care, so that their 
autonomy does not get impaired. Robots are far from being able to replace human 
caregivers 
altogether, not to mention that for most situations this is likely an 
undesirable 
goal. Therefore, the teaming of humans and robots in assistance-giving is the 
objective we are proposing. Robots can help structure assistance giving 
interactions between caregivers and care recipients. An example that doesn’t 
come 
from the health care context, but from teaching, illustrates some possible 
functions for the robot: providing instructions for the task, assigning roles, 
and prompting the caregiver to offer different types of input that could be 
corrective feedback, praise, encouragement etc. Chandra et al. \cite{chandra2015can} compared 
a robot and a human facilitator of a collaborative learning activity. Children 
engaged in a learning-by-teaching task, in which one child taught the other how 
to write different letters or words. Either a robot or a human acted as 
facilitators by introducing the task, assigning roles (teacher or learner), 
providing instruction throughout the task and prompting the teacher-child to 
provide corrective feedback to the learner child. The video and audio 
recordings of the session were coded. Teacher-children provided more extended 
corrective feedback with the robot facilitator and more minimal corrective 
feedback with the human facilitator. Authors argue that the 
teacher-children felt more responsible regarding their performance in the 
presence of the robot. Combining these results with the duration of gaze that 
the facilitator directed towards the children (the robot made longer-duration 
gazes than the human facilitator) the authors conclude that two different 
patterns of interpersonal distancing emerged: in the case of the robot 
facilitator children followed the reciprocity model (responding to closeness 
with closeness), in the case of the human facilitator they followed the 
compensation model (responding to distancing with closeness).

The overall goal of having structured interactions is to ensure that they are 
meaningful, positive and inclusive. This is beneficial for the strengthening of 
relationships between people with health care conditions and health care 
providers, caregivers, and others. Most importantly, these robot-assisted 
interactions 
should improve the quality of life and sense of well-being of the person with 
the health-condition. This is why one of the functions of SARs needs to be that 
of engendering positive feelings for people with health conditions in social 
contexts.

\subsection{Changing how a person with chronic illness feels in a social 
context}
\label{sec:2.5}

Social situations can be stressful for people with health conditions. This can 
be due to the specifics of the health condition, for example, people with 
Post-Traumatic Stress Disorder
can feel uncomfortable in social situations that trigger traumatic memories 
\cite{nietlisbach2009social}, but more generally it can be caused by the stigma 
associated with 
health conditions \cite{weiss2006health}. Stigma can take various forms: 
feeling ostracized, 
devalued, scorned \cite{dovidio2000stigma}. Many people with health conditions 
experience 
psychological distress from perceived stigma from others 
\cite{van2006measuring}. 

\subsubsection{Promoting positive feelings in interactions}
\label{sec:2.5.1}

We’ve already discussed studies of robots that can help people experience more 
positive feelings in social interactions. These ideas can be used to create 
SARs that help combat some of the negative effects of stigma. For example, 
behaviors of the robot used by Tennent, Shen and Jung (2019), such as inviting 
people to join a conversation through movement, could be used for 
developing and testing robots that help people with health conditions feel 
welcomed and encouraged to participate in social interactions. Also, behaviors 
from the robot used by Mutlu et al. \cite{mutlu2009footing}, such as the use of 
gaze to suggest conversation roles, could be adapted to create feelings 
of inclusiveness for people with health conditions. 

An example of a robot 
specifically designed for influencing how a person feels in a social 
interaction with another human was tested by Pettinati, Arkin and Shim \cite{pettinati2016influence}. 
They used a social robot (Nao) for active listening. The robot was envisioned 
as a peripheral addition to an interaction between two people. The robot 
indicated active listening by turning its head towards the person speaking. 
Participants perceived the active robot as having more of a social presence 
than the controls (a non-active Nao and a plush toy) but participants felt 
equally comfortable self-disclosing in front of the active robot. The lack of a 
negative impact of the robot’s presence for self-disclosure is encouraging for 
the prospects of designing a mediator robot that does not detract from the 
interaction between humans. The absence of negative effects is a start, but 
further research is needed to establish whether the robot contributed any 
additional positive psychological effects of feeling listened to when 
disclosing personal information to another person. 

\subsubsection{Mitigating negative feelings in interactions}
\label{sec:2.5.2}
In this paper we 
specifically review studies that used robots to support social interactions 
between people, but many ideas from human-robot interaction studies can be 
adapted to the social mediation context. For example, roboticists are 
developing pet-like 
robots to assist with stress reduction during counseling sessions 
\cite{bethel2016using}. Stress-reducing robots could also be used to help 
people with social anxiety in a variety of social circumstances.

Although still in its initial stages, the development of mediator SARs for the 
social management of health is replete with opportunities for further design 
and HRI research. However, challenges of designing, testing and beneficially 
integrating these systems into our lives and health management also warrant 
discussion.

\section{Challenges of designing and using mediator SARs}
\label{sec:3}

There are four classes of challenges that exist with regards to designing and 
using SARs for the social management of health: a) challenges related to the 
status and well-being of the person with the health condition, b) challenges 
related to the 
impact of SARs on human-human interactions, especially the unforeseen or 
unwanted effects,  c) challenges related to the broader social 
and cultural context and d) challenges related to the features and usefulness 
of the robot itself. For a successful embedding of SARs in the caregiving 
context, these challenges will need to be overcome through ingenious design and 
most importantly careful research.

\subsection{Challenges related to the status and well-being of the person with 
the health condition}
\label{sec:3.1}

\subsubsection{Preservation of autonomy and dignity}
\label{sec:3.1.1}
In a mediator role, SARs will assist interactions between two or more people. 
However, the health and well-being of the person with the health condition 
using the SAR 
is of primary importance, as this is the reason for developing SARs in the 
first place. The challenge with giving any type of assistance (but perhaps even 
more importantly when giving assistance through the use robots) is the 
preservation of the person’s autonomy and dignity. Sharkey and Sharkey 
\cite{sharkey2012granny},
warned that careless use of assistive robots could lead to a loss of control of 
important aspects of one’s life and feelings of objectification, and Wilson et 
al. \cite{wilson2016autonomy} proppose that the concepts of autonomy and 
personal dignity, which are 
guiding ethical principles in occupational therapy, should be incorporated into 
the desgin process of social robots.
Because people 
with health conditions are a vulnerable population, there is concern that 
robotic assistance would lead to a loss of personal liberty. 
One way in which 
this could happen is through overreliance on the robot, leading to enfeeblement 
and then dependence. If the robot completely takes over a certain task or 
important aspects of it (with regards to the robot functions proposed by this 
paper, one such task is the 
management of interactions) the worry is that people might lose the ability to 
perform the task themselves. For example, if a person becomes overly reliant on 
the robot alerting them to problematic nonveral aspects of a conversation (a 
function explored in Section \ref{sec:2.3.1}) instead of using the robot's 
feedback  to 
improve one's attention to cues from the interlocutor, this might lead to 
more problematic interactions in the future when the robot is not present.   
With some tasks this might be fine, as the person 
might have already lost that ability because of the health condition (for 
example, for severe dementia the function of redirecting conversation to 
non-repetitive topics might be needed for the remainder of the person's care), 
but with 
others, effortful attempts to maintain abilities might be desirable for 
independence. SARs involved in the social management of health should thus 
support rather than take over the task of initiating and sustaining 
interactions between people. As mentioned above, the right level of direction 
and assistance should be established through research.

\subsubsection{Ownership, control and authority of the SAR}
\label{sec:3.1.2}

Another way in which personal liberty of people could be encroached on has to 
do with the status of the person with the health care condition with regards to 
the SAR: who owns the SAR and who controls it? \cite{arnold2017beyond} Also, 
what obligations does that SAR have towards the different people that are part 
of the caregiving ecosystem? \cite{wilson2016reflections} This is an 
especially important 
consideration for the SAR functions that we propose in this paper. We are 
focusing on robots that can manage social interactions between people, and 
although the ultimate goal of the robot is to support the social management of 
health of the person with the health condition, precisely because it is a robot 
designed for supporting interactions between humans, the robot would serve 
multiple people, including health care providers, caregivers and other 
people belonging to the social circle of the person with the health condition. 
Also, given that health conditions can impair people’s judgement, it is not 
always feasible that the authority over the robot and its use remains with the 
person with the health condition. In fact, in some situations it might be 
desirable that the robot itself exert authority over the person with the health 
condition. We learned from the study by Shim, Arkin and Pettinati 
\cite{shim2017intervening} that 
people felt that the robot should never have the authority to judge patients. 
On the other hand, participants in the Utami and Bickmore study 
\cite{utami2019collaborative} welcomed the 
mild social pressure from the robot when the robot successfully prompted them 
to perform the 
therapeutic interaction exercises. Even more so, caregivers participating in 
the study by Moharana et al. \cite{moharana2019robots} wanted a robot to have 
much more authority and adopt
the role of a neutral third party who would determine the person receiving care 
to do 
things that they do not wish to do, but need to for their own good, for 
example, 
taking their medication. Some participants even envisioned that the robot would 
do this using the doctor’s voice. The balance between assistance and autonomy 
should be decided preferably on a case by case basis and by taking into account 
the context. However, the functions we specify in this paper are very much 
subservient to the goals they try to achieve, which is not just preventing 
isolation, but also preserving autonomy and preventing dehumanization and 
stigma.

\subsubsection{Deception and unidirectional emotional bonds}
\label{sec:3.1.3}

Another aspect of using SARs that has been flagged as potentially contributing 
negatively to the life and dignity of the person assisted is the issue of 
deception \cite{moharana2019robots}, infantilization \cite{sharkey2012granny}  
and inauthenticity of the human-robot interaction \cite{turkle2006relational}. 
SARs 
capitalize on the deeply ingrained human propensity to engage with lifelike 
social behavior and use this engagement for natural interaction with people 
\cite{okamura2010medical}.

Robots today can behave in lifelike, social ways, but they are neither alive 
nor do they actually feel any social emotions. But the person assisted by the 
robot, especially those who are struggling with cognitive impairments, can be 
tricked (much like children are), by the robot’s behavior into believing the 
robot is something it is not. Especially when features such as touch (which 
would very likely be available in a healthcare robot) may amplify feelings of 
intimacy \cite{arnold2017tactile}. This could lead to the formation of 
unidirectional emotional bonds in which the person harbors feeling for the 
robot but the 
robot is ontologically unable to reciprocate \cite{scheutz201113}. This could 
be particularly 
problematic when the SAR is used for long periods of time and attachment is 
developed. As Sharkey and Sharkey, discuss, 
there are 
different levels of “buying into” the robot’s behavior and acting 
“as if” the robot truly had social feelings, some of which are acceptable and 
some which border ethical concern. The functions we envision for SARs in this 
paper, namely that of supporting social interactions, could perhaps mitigate 
some of the concerns regarding deception and formation of problematic emotional 
bonds. In its most offending form, deception 
from SARs is when people start believing that the SAR is a companion that 
understands and shares their deepest feelings. The functions we propose for 
SARs shift the focus from the human-robot relationships to the human-human 
relationships, for which the robot simply offers support. The purpose of the 
robot 
intervening is not for it to offer companionship, but to optimize the ways in 
which people offer companionship to each other. Additionally, having another 
human in the loop (often the caregiver), can help with the supervision and 
correction of any problematic aspects of the relationship between the robot and 
the person assisted.

\subsection{Challenges related to the impact of SARs on human-human 
interactions}
\label{sec:3.2}

\subsubsection{Potential reduction in human contact}
\label{sec:3.2.1}

With regards to human-human interactions, a common concern raised in relation 
to SARs in general is the potential drastic reduction in human contact 
\cite{sharkey2012granny,moharana2019robots}. If caregiving tasks are taken 
over by 
robots, the fear is that humans needing assistance will end up interacting 
mostly with robots rather than other fellow humans, and this will have 
detrimental effects on their social life and health. This concern is especially 
pertinent to the function of SARs as providers of companionship. However, the 
vision presented in this paper, is quite the opposite. We suggest that SARs 
should adopt mediator roles and assist people with health conditions in their 
social management of health. We propose not for robots to diminish or replace 
human social contact, but on the contrary, to increase and enhance it. This 
paper thus proposes functions for SARs that are different from the ones 
evaluated 
by Sharkey and Sharkey, which focused on SARs assisting with daily tasks, 
monitoring behavior and health and providing companionship. 

\subsubsection{Alteration of human-human interactions}
\label{sec:3.2.2}

However, our vision is subject to a different concern: that mediator robots 
would inadvertently alter and negatively impact human-human interactions. A 
robot embedded in an interaction could detract from it by being an unwelcomed 
distraction \cite{tennent2019micbot}. Instead of focusing on each other, 
people would instead focus on the robot and change their interaction to 
accommodate the robot. A way to think about this issue is in terms of 
foregrounding or backgrounding of interactions by robots, and the amount of 
direction they offer \cite{moharana2019robots}. Based on the specific needs of the 
interaction and of the interactants, the robot could take a peripheral role, 
subtly cueing people to potential opportunities or problems in their 
interactions, or a more leading role, directing the interaction between people. 
Moharana et al. suggest for example that in the early stages of dementia, and 
when the interaction is positive and satisfying for both the caregiver and 
the person receiving care, a mediator SAR could have a peripheral role in 
interactions. 
However, as the disease progresses and interactions become more frustrating, 
for example, because of agitation and forgetfulness, the robot could take on 
more the role of conversation partner in the interaction, taking over the 
stressful task of answering repetitive questions and providing redirection. 
However, it 
is important for the robot to not only intervene in negative situations, 
but also when it detects opportunities for positive social interactions, lest 
it be perceived as a ``watchdog'' and its interventions associated with 
unpleasant events \cite{shen2018stop}. 
In the sections above, we’ve 
seen examples of mediation from both peripheral robotic devices, such as the 
ones from Hoffman et al. \cite{hoffman2015design} and Tennent, Shen and Jung \cite{tennent2019micbot}, and also 
mediation 
from robots in leading roles, offering high amounts of direction such as those 
developed by Shen, Slovak and Jung \cite{shen2018stop} or Utami and Bickmore \cite{utami2019collaborative}. Further 
research is needed to establish the factors that should dictate the degree of 
robot involvement in an interaction. The factors proposed by Moharana et 
al., namely stage 
of health condition and positivity of interaction, are a good start, but more 
factors need to be tested, including but not limited to the preference and 
personality of the interactants or the type of interaction.

\subsubsection{Disruption of intimacy and privacy of interactions}
\label{sec:3.2.3}

SARs, through their social presentence could also disturb the intimacy and 
privacy \cite{sharkey2012granny} of the interaction and actualize the 
proverbial ``two is company, three is a crowd''. As we’ve seen, Pettinati, 
Arkin 
and Shim \cite{pettinati2016influence} found promisingly that the robot’s presence did not have any 
negative effects on self-disclosure when embedded in an interaction between two 
people, however more research is needed to establish that this is the case 
across contexts. Pettinati et al. only showed this in the context of a 
conversation between two strangers, an interviewer and an interviewee, not 
between, for example, people who know each other and have a long relationship 
history. On the 
other hand, the robot’s presence might in some cases be more tolerable than 
that of another person. Participants in the couple’s therapy study by Utami and 
Bickmore \cite{utami2019collaborative} indicated that it was easier for them to 
perform the exercises 
and disclose things in front of the robot than it would have been in front of 
a human therapist. More generally, Mutlu et al. \cite{mutlu2009footing} showed that robots can have an 
effect on how people feel about an interaction. Of course, this possibility is 
a great opportunity to use the robot’s leverage to create positive interactions 
between people, but it is also a warning sign that unintended negative effects 
might also occur, and they should be carefully researched.

\subsection{Challenges related to the broader social and cultural context}
\label{sec:3.3}

The caregivers and the care recipients assisted by the robot are not the 
only 
ones that need 
to be considered in designing the SAR. It is important that the robot is 
seamlessly embedded in the social and cultural context. Cultural differences 
exist with regards to caregiving and illness \cite{moharana2019robots} which 
result in different roles, degrees of autonomy, and experiences for the 
caregiver and the person being cared for. Also, different cultures may have 
different 
attitudes towards robots, their form and functions \cite{bartneck2005cultural}. 
An example of how to 
ensure the robot fits the needs of the community it serves, is the study by 
Joshi and Šabanović \cite{joshi2019robots}, which worked with the local community to better 
understand their goals in terms of integrating robots in the context of 
social interactions. For example, prior to designing the activities and 
introducing the robots, Joshi and Šabanović, 
conducted extensive interviews with the staff at the preschool and the 
assistive living-dementia care center where the robots would be used. The 
interviews helped them identify the following community goal: to engage older 
adults and children in 
activities that 
were meaningful for both groups, with the purpose of facilitating relations 
similar to 
grandparents 
and grandchildren. The authors then systematically investigated the 
usefulenes of different robots for achieving this goal. They conclude that 
some robots were not  well suited for what that community wanted. For example 
the Cozmo robot led to activities that were too fast-paced for the older 
adults, 
and which distracted the children from meaningful intergenerational engagement 
rather than facilitating interaction.

\subsection{Challenges related to the features and usefulness of the SAR}
\label{sec:3.4}

\subsubsection{Ability to adapt}
\label{sec:3.4.1}

A key challenge and feature of the SAR, in order for it to be successful, will 
be its ability to adapt  \cite{moharana2019robots,okamura2010medical}. 
Adaptability is important to keep pace with the progression of the health 
condition and the changing needs and contexts of the person assisted. In many 
of the studies discussed, the positive effect of the robot on social 
interactions stems from the robot being an interesting gadget that prompted 
people to interact with each other about it. However, we know little about what 
would happen once the novelty effect wears off. Ideally, the robot and its 
repertoire of interventions would continue to change over time both as 
technology progresses and as more research establishes new effective 
interventions. The SAR should also be personalized to the preferences and needs 
of the person using it \cite{moharana2019robots,okamura2010medical}. People 
react differently to different intervention styles. A major gap in the 
literature describing uses of robots as mediators of human-human interactions, 
is the lack of studies focusing on individual differences and how they modulate 
the robot’s effect.

\subsubsection{Creation and meeting of expectations}
\label{sec:3.4.2}

Connected to the challenge of deception explored above, SARs should be designed 
in mindful ways that do not create expectations that are not met \cite{tennent2019micbot}. For example, just because a robot can offer suggestions of 
conversation topics, it does not mean that it has an understanding of what 
people talk about. The status of the mediator robot as something in between a 
tool and a social interaction partner needs to be given proper consideration. 
As mentioned above, features that subconsciously convey social signals and 
imply capabilities that the SAR does not have (such as touch conveying social 
bonding and a capability for affection) should be carefully researched before 
being implemented. Roboticists should also be mindful about expectations 
regarding 
avaiability of the SAR. As discussed above, the SAR should not lead to 
enfeeblement and loss of autonomy.  

\subsubsection{Robustness and safety}
\label{sec:3.4.3}

Finally, SARs need to be robust in terms of their ability to carry out the 
functions they are designed for. Since SARs for the social management of health 
are envisioned to assist vulnerable populations, potential technical problems 
need to be reduced to a minimum \cite{tennent2019micbot}. When robots 
that simply provide entertainment fail, the failure might be more tolerable and 
less costly, but when people rely on robots for tasks that have significance 
for their health, technical issues become seriously problematic.

\begin{figure*}
\vskip-8mm
\hglue26mm
\centerline{\includegraphics[width=1.15\textwidth]{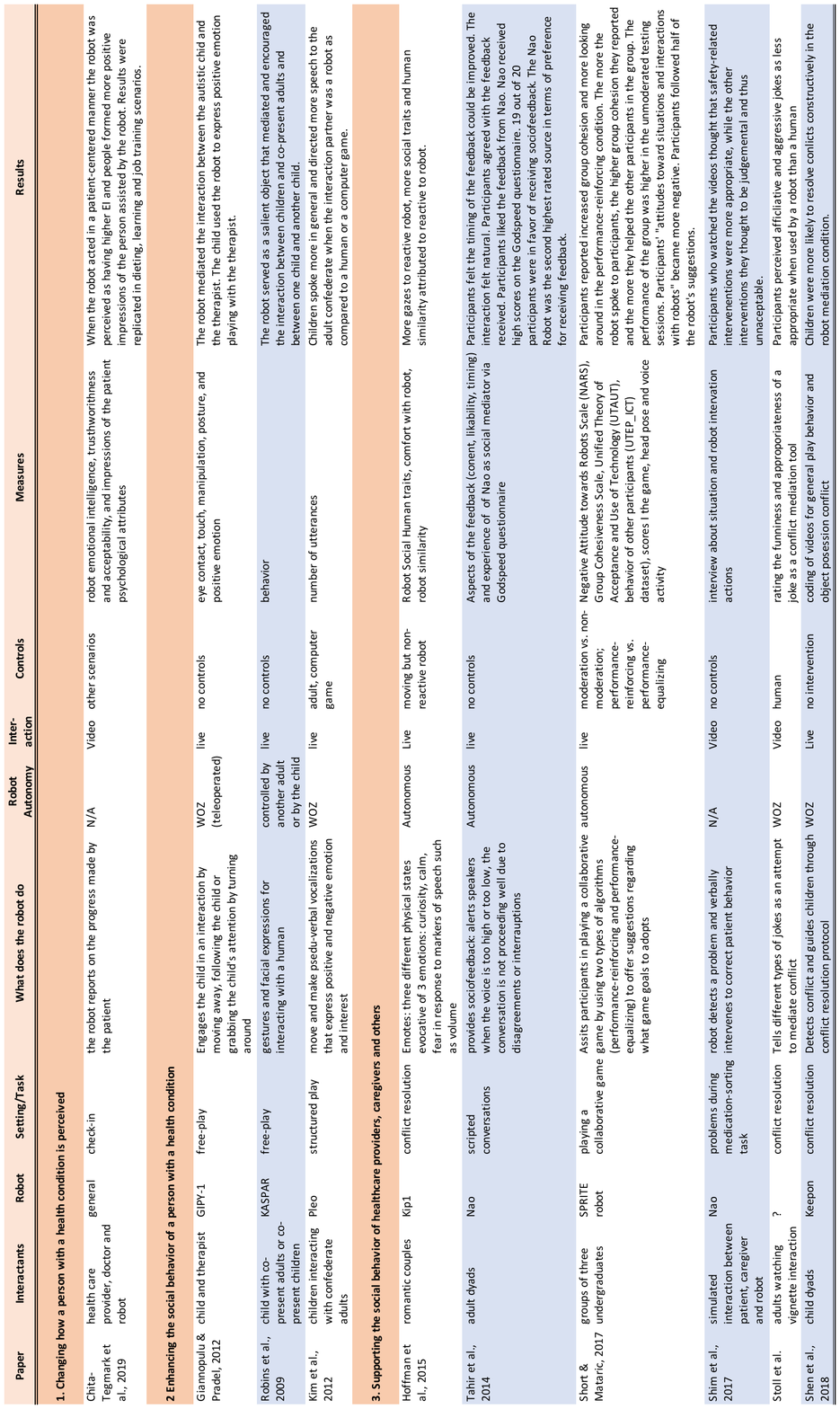}}
\caption{HRI studies of robots mediating social interactions between people 
(continued on the next page)}
\label{fig:2a}       % Give a unique label
\end{figure*}
%

\begin{figure*}
\vskip-12mm
\hglue3mm
\centerline{\includegraphics[width=1.15\textwidth]{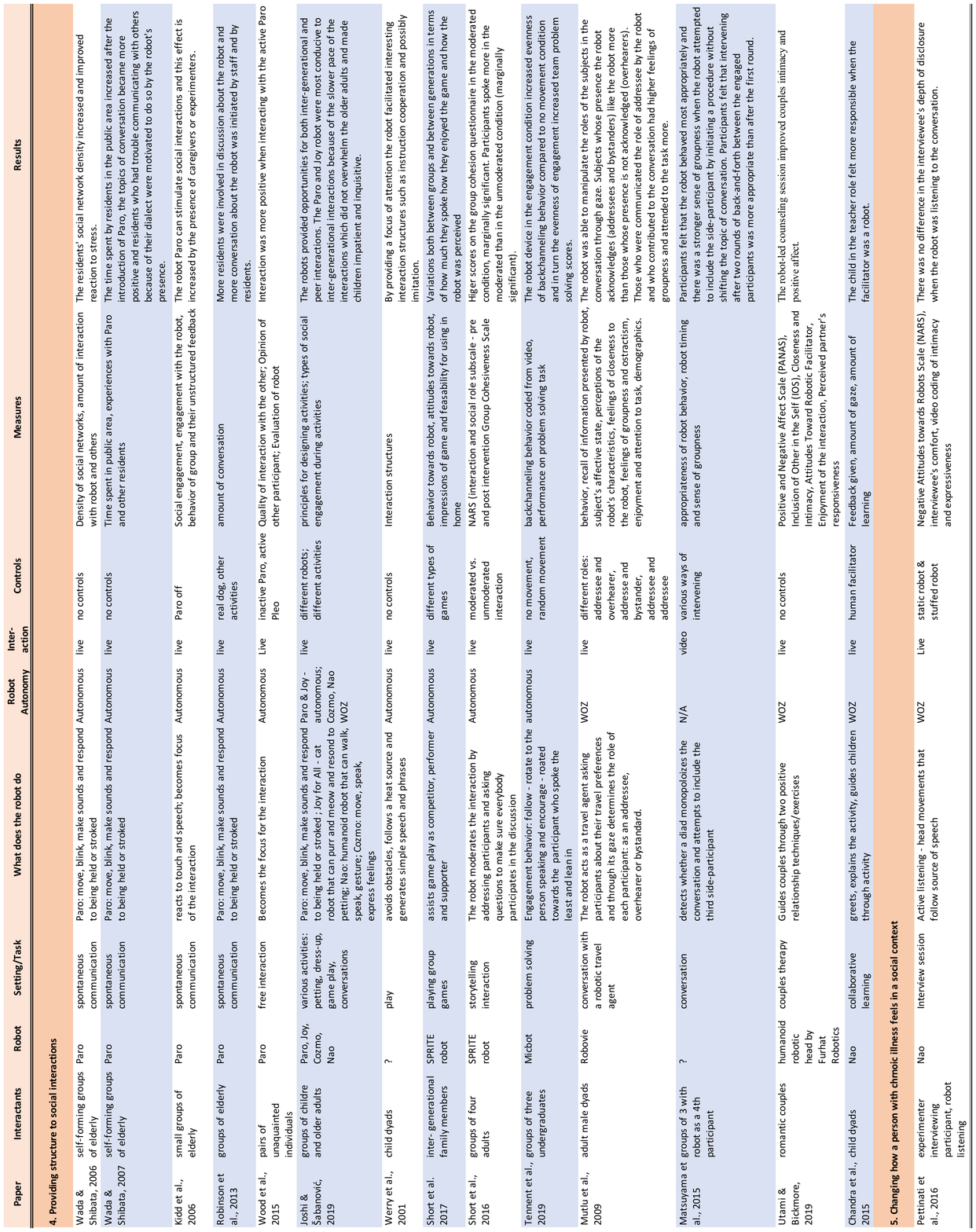}}
	\label{fig:2b}       % Give a unique label
\end{figure*}

\section{Conclusion}
\label{sec:4}

In this paper we proposed five classes of functions for SARs that would support 
the social management of health by assisting human-human interactions. We’ve 
identified the research gaps in our understanding of how a robot could change 
the way a person with a health condition is perceived by others. We have 
illustrated through some 
previous results, mainly from case studies, how robots could enhance the 
social behavior of people with health conditions by addressing the impairments 
specific to the health condition. We summarized the research on how robots can 
modify the social behavior of people both for further enhancing positive 
interactions and for correcting negative ones. We surveyed the research 
studies that have used various levels of robot intervention to structure 
human-human interactions in both clinical and non-clinical settings. 
Finally, we exemplified through previous findings how people’s feelings in 
a social 
context 
might be changed for the better by the introduction of a robot into the 
interaction. 
While reviewing the literature relevant for the mediator role for 
SARs, we have identified opportunities for further research and robot design. 
We discussed potential challenges in the design and use of SARs and 
showed that when the focus of the SAR's intervention is on the enhancement of 
the human-human interaction not on the replacement of caregivers, many of the 
general concerns with regards to SARs can be mitigated. The existing literature 
and the promising research avenues identified suggest that the development of 
SARs for the management of social interactions could yield important benefits 
for health.
\\
\\

\textbf{Funding: }This project was supported by the National Science Foundation 
grant IIS-1316809.

\textbf{Conflict of interest: }The authors declare that they have no conflict 
of 
interest.

%\paragraph{Paragraph headings} Use paragraph headings as needed.
%\begin{equation}
%\end{equation}

\bibliography{mediatorrobotreview}

\begin{thebibliography}{85}%
\makeatletter
\providecommand \@ifxundefined [1]{%
 \@ifx{#1\undefined}
}%
\providecommand \@ifnum [1]{%
 \ifnum #1\expandafter \@firstoftwo
 \else \expandafter \@secondoftwo
 \fi
}%
\providecommand \@ifx [1]{%
 \ifx #1\expandafter \@firstoftwo
 \else \expandafter \@secondoftwo
 \fi
}%
\providecommand \natexlab [1]{#1}%
\providecommand \enquote  [1]{``#1''}%
\providecommand \bibnamefont  [1]{#1}%
\providecommand \bibfnamefont [1]{#1}%
\providecommand \citenamefont [1]{#1}%
\providecommand \href@noop [0]{\@secondoftwo}%
\providecommand \href [0]{\begingroup \@sanitize@url \@href}%
\providecommand \@href[1]{\@@startlink{#1}\@@href}%
\providecommand \@@href[1]{\endgroup#1\@@endlink}%
\providecommand \@sanitize@url [0]{\catcode `\\12\catcode `\$12\catcode
  `\&12\catcode `\#12\catcode `\^12\catcode `\_12\catcode `\%12\relax}%
\providecommand \@@startlink[1]{}%
\providecommand \@@endlink[0]{}%
\providecommand \url  [0]{\begingroup\@sanitize@url \@url }%
\providecommand \@url [1]{\endgroup\@href {#1}{\urlprefix }}%
\providecommand \urlprefix  [0]{URL }%
\providecommand \Eprint [0]{\href }%
\providecommand \doibase [0]{http://dx.doi.org/}%
\providecommand \selectlanguage [0]{\@gobble}%
\providecommand \bibinfo  [0]{\@secondoftwo}%
\providecommand \bibfield  [0]{\@secondoftwo}%
\providecommand \translation [1]{[#1]}%
\providecommand \BibitemOpen [0]{}%
\providecommand \bibitemStop [0]{}%
\providecommand \bibitemNoStop [0]{.\EOS\space}%
\providecommand \EOS [0]{\spacefactor3000\relax}%
\providecommand \BibitemShut  [1]{\csname bibitem#1\endcsname}%
\let\auto@bib@innerbib\@empty
%</preamble>
\bibitem [{\citenamefont {Cohen}(2004)}]{cohen2004social}%
  \BibitemOpen
  \bibfield  {author} {\bibinfo {author} {\bibfnamefont {Sheldon}\ \bibnamefont
  {Cohen}},\ }\bibfield  {title} {\enquote {\bibinfo {title} {Social
  relationships and health.}}\ }\href@noop {} {\bibfield  {journal} {\bibinfo
  {journal} {American psychologist}\ }\textbf {\bibinfo {volume} {59}},\
  \bibinfo {pages} {676} (\bibinfo {year} {2004})}\BibitemShut {NoStop}%
\bibitem [{\citenamefont {Umberson}\ and\ \citenamefont
  {Karas~Montez}(2010)}]{umberson2010social}%
  \BibitemOpen
  \bibfield  {author} {\bibinfo {author} {\bibfnamefont {Debra}\ \bibnamefont
  {Umberson}}\ and\ \bibinfo {author} {\bibfnamefont {Jennifer}\ \bibnamefont
  {Karas~Montez}},\ }\bibfield  {title} {\enquote {\bibinfo {title} {Social
  relationships and health: A flashpoint for health policy},}\ }\href@noop {}
  {\bibfield  {journal} {\bibinfo  {journal} {Journal of health and social
  behavior}\ }\textbf {\bibinfo {volume} {51}},\ \bibinfo {pages} {S54--S66}
  (\bibinfo {year} {2010})}\BibitemShut {NoStop}%
\bibitem [{\citenamefont {de~Jong~Gierveld}\ \emph {et~al.}(2006)\citenamefont
  {de~Jong~Gierveld}, \citenamefont {Van~Tilburg},\ and\ \citenamefont
  {Dykstra}}]{dejong2006loneliness}%
  \BibitemOpen
  \bibfield  {author} {\bibinfo {author} {\bibfnamefont {Jenny}\ \bibnamefont
  {de~Jong~Gierveld}}, \bibinfo {author} {\bibfnamefont {Theo}\ \bibnamefont
  {Van~Tilburg}}, \ and\ \bibinfo {author} {\bibfnamefont {Pearl~A}\
  \bibnamefont {Dykstra}},\ }\bibfield  {title} {\enquote {\bibinfo {title}
  {Loneliness and social isolation},}\ }\href@noop {} {\bibfield  {journal}
  {\bibinfo  {journal} {Cambridge handbook of personal relationships}\ ,\
  \bibinfo {pages} {485--500}} (\bibinfo {year} {2006})}\BibitemShut {NoStop}%
\bibitem [{\citenamefont {Cacioppo}\ and\ \citenamefont
  {Cacioppo}(2014)}]{cacioppo2014social}%
  \BibitemOpen
  \bibfield  {author} {\bibinfo {author} {\bibfnamefont {John~T}\ \bibnamefont
  {Cacioppo}}\ and\ \bibinfo {author} {\bibfnamefont {Stephanie}\ \bibnamefont
  {Cacioppo}},\ }\bibfield  {title} {\enquote {\bibinfo {title} {Social
  relationships and health: The toxic effects of perceived social isolation},}\
  }\href@noop {} {\bibfield  {journal} {\bibinfo  {journal} {Social and
  personality psychology compass}\ }\textbf {\bibinfo {volume} {8}},\ \bibinfo
  {pages} {58--72} (\bibinfo {year} {2014})}\BibitemShut {NoStop}%
\bibitem [{\citenamefont {Tickle-Degnen}\ \emph {et~al.}(2014)\citenamefont
  {Tickle-Degnen}, \citenamefont {Scheutz},\ and\ \citenamefont
  {Arkin}}]{tickle2014collaborative}%
  \BibitemOpen
  \bibfield  {author} {\bibinfo {author} {\bibfnamefont {Linda}\ \bibnamefont
  {Tickle-Degnen}}, \bibinfo {author} {\bibfnamefont {Matthias}\ \bibnamefont
  {Scheutz}}, \ and\ \bibinfo {author} {\bibfnamefont {Ronald~C}\ \bibnamefont
  {Arkin}},\ }\bibfield  {title} {\enquote {\bibinfo {title} {Collaborative
  robots in rehabilitation for social self-management of health},}\ }\href@noop
  {} {\  (\bibinfo {year} {2014})}\BibitemShut {NoStop}%
\bibitem [{\citenamefont {Okamura}\ \emph {et~al.}(2010)\citenamefont
  {Okamura}, \citenamefont {Mataric},\ and\ \citenamefont
  {Christensen}}]{okamura2010medical}%
  \BibitemOpen
  \bibfield  {author} {\bibinfo {author} {\bibfnamefont {Allison~M}\
  \bibnamefont {Okamura}}, \bibinfo {author} {\bibfnamefont {Maja~J}\
  \bibnamefont {Mataric}}, \ and\ \bibinfo {author} {\bibfnamefont {Henrik~I}\
  \bibnamefont {Christensen}},\ }\bibfield  {title} {\enquote {\bibinfo {title}
  {Medical and health-care robotics},}\ }\href@noop {} {\bibfield  {journal}
  {\bibinfo  {journal} {IEEE Robotics \& Automation Magazine}\ }\textbf
  {\bibinfo {volume} {17}},\ \bibinfo {pages} {26--37} (\bibinfo {year}
  {2010})}\BibitemShut {NoStop}%
\bibitem [{\citenamefont {Rabbitt}\ \emph {et~al.}(2015)\citenamefont
  {Rabbitt}, \citenamefont {Kazdin},\ and\ \citenamefont
  {Scassellati}}]{rabbitt2015integrating}%
  \BibitemOpen
  \bibfield  {author} {\bibinfo {author} {\bibfnamefont {Sarah~M}\ \bibnamefont
  {Rabbitt}}, \bibinfo {author} {\bibfnamefont {Alan~E}\ \bibnamefont
  {Kazdin}}, \ and\ \bibinfo {author} {\bibfnamefont {Brian}\ \bibnamefont
  {Scassellati}},\ }\bibfield  {title} {\enquote {\bibinfo {title} {Integrating
  socially assistive robotics into mental healthcare interventions:
  Applications and recommendations for expanded use},}\ }\href@noop {}
  {\bibfield  {journal} {\bibinfo  {journal} {Clinical psychology review}\
  }\textbf {\bibinfo {volume} {35}},\ \bibinfo {pages} {35--46} (\bibinfo
  {year} {2015})}\BibitemShut {NoStop}%
\bibitem [{\citenamefont {Van~der Putte}\ \emph {et~al.}(2019)\citenamefont
  {Van~der Putte}, \citenamefont {Boumans}, \citenamefont {Neerincx},
  \citenamefont {Rikkert},\ and\ \citenamefont {de~Mul}}]{van2019social}%
  \BibitemOpen
  \bibfield  {author} {\bibinfo {author} {\bibfnamefont {Daisy}\ \bibnamefont
  {Van~der Putte}}, \bibinfo {author} {\bibfnamefont {Roel}\ \bibnamefont
  {Boumans}}, \bibinfo {author} {\bibfnamefont {Mark}\ \bibnamefont
  {Neerincx}}, \bibinfo {author} {\bibfnamefont {Marcel~Olde}\ \bibnamefont
  {Rikkert}}, \ and\ \bibinfo {author} {\bibfnamefont {Marleen}\ \bibnamefont
  {de~Mul}},\ }\bibfield  {title} {\enquote {\bibinfo {title} {A social robot
  for autonomous health data acquisition among hospitalized patients: An
  exploratory field study},}\ }in\ \href@noop {} {\emph {\bibinfo {booktitle}
  {2019 14th ACM/IEEE International Conference on Human-Robot Interaction
  (HRI)}}}\ (\bibinfo {organization} {IEEE},\ \bibinfo {year} {2019})\ pp.\
  \bibinfo {pages} {658--659}\BibitemShut {NoStop}%
\bibitem [{\citenamefont {Briggs}\ \emph {et~al.}(2015)\citenamefont {Briggs},
  \citenamefont {Scheutz},\ and\ \citenamefont
  {Tickle-Degnen}}]{briggs2015robots}%
  \BibitemOpen
  \bibfield  {author} {\bibinfo {author} {\bibfnamefont {Priscilla}\
  \bibnamefont {Briggs}}, \bibinfo {author} {\bibfnamefont {Matthias}\
  \bibnamefont {Scheutz}}, \ and\ \bibinfo {author} {\bibfnamefont {Linda}\
  \bibnamefont {Tickle-Degnen}},\ }\bibfield  {title} {\enquote {\bibinfo
  {title} {Are robots ready for administering health status surveys': First
  results from an hri study with subjects with parkinson's disease},}\ }in\
  \href@noop {} {\emph {\bibinfo {booktitle} {Proceedings of the Tenth Annual
  ACM/IEEE International Conference on Human-Robot Interaction}}}\ (\bibinfo
  {organization} {ACM},\ \bibinfo {year} {2015})\ pp.\ \bibinfo {pages}
  {327--334}\BibitemShut {NoStop}%
\bibitem [{\citenamefont {Wilson}\ \emph
  {et~al.}(2016{\natexlab{a}})\citenamefont {Wilson}, \citenamefont
  {Tickle-Degnen},\ and\ \citenamefont {Scheutz}}]{wilson2016designing}%
  \BibitemOpen
  \bibfield  {author} {\bibinfo {author} {\bibfnamefont {Jason~R}\ \bibnamefont
  {Wilson}}, \bibinfo {author} {\bibfnamefont {Linda}\ \bibnamefont
  {Tickle-Degnen}}, \ and\ \bibinfo {author} {\bibfnamefont {Matthias}\
  \bibnamefont {Scheutz}},\ }\bibfield  {title} {\enquote {\bibinfo {title}
  {Designing a social robot to assist in medication sorting},}\ }in\ \href@noop
  {} {\emph {\bibinfo {booktitle} {International Conference on Social
  Robotics}}}\ (\bibinfo {organization} {Springer},\ \bibinfo {year} {2016})\
  pp.\ \bibinfo {pages} {211--221}\BibitemShut {NoStop}%
\bibitem [{\citenamefont {Kim}\ \emph {et~al.}(2013{\natexlab{a}})\citenamefont
  {Kim}, \citenamefont {Jeon}, \citenamefont {Im}, \citenamefont {Seo},
  \citenamefont {Cho}, \citenamefont {Noh}, \citenamefont {Yoon}, \citenamefont
  {Kim}, \citenamefont {Ye}, \citenamefont {Chin} \emph
  {et~al.}}]{kim2013structural}%
  \BibitemOpen
  \bibfield  {author} {\bibinfo {author} {\bibfnamefont {Geon~Ha}\ \bibnamefont
  {Kim}}, \bibinfo {author} {\bibfnamefont {Seun}\ \bibnamefont {Jeon}},
  \bibinfo {author} {\bibfnamefont {Kiho}\ \bibnamefont {Im}}, \bibinfo
  {author} {\bibfnamefont {Sang~Won}\ \bibnamefont {Seo}}, \bibinfo {author}
  {\bibfnamefont {Hanna}\ \bibnamefont {Cho}}, \bibinfo {author} {\bibfnamefont
  {Young}\ \bibnamefont {Noh}}, \bibinfo {author} {\bibfnamefont {Cindy}\
  \bibnamefont {Yoon}}, \bibinfo {author} {\bibfnamefont {Hee-Jin}\
  \bibnamefont {Kim}}, \bibinfo {author} {\bibfnamefont {Byoung~Seok}\
  \bibnamefont {Ye}}, \bibinfo {author} {\bibfnamefont {Ju~Hee}\ \bibnamefont
  {Chin}},  \emph {et~al.},\ }\bibfield  {title} {\enquote {\bibinfo {title}
  {Structural brain changes after robot-assisted cognitive training in the
  elderly: A single-blind randomized controlled trial},}\ }\href@noop {}
  {\bibfield  {journal} {\bibinfo  {journal} {Alzheimer's \& Dementia: The
  Journal of the Alzheimer's Association}\ }\textbf {\bibinfo {volume} {9}},\
  \bibinfo {pages} {P476--P477} (\bibinfo {year}
  {2013}{\natexlab{a}})}\BibitemShut {NoStop}%
\bibitem [{\citenamefont {Banks}\ \emph {et~al.}(2008)\citenamefont {Banks},
  \citenamefont {Willoughby},\ and\ \citenamefont {Banks}}]{banks2008animal}%
  \BibitemOpen
  \bibfield  {author} {\bibinfo {author} {\bibfnamefont {Marian~R}\
  \bibnamefont {Banks}}, \bibinfo {author} {\bibfnamefont {Lisa~M}\
  \bibnamefont {Willoughby}}, \ and\ \bibinfo {author} {\bibfnamefont
  {William~A}\ \bibnamefont {Banks}},\ }\bibfield  {title} {\enquote {\bibinfo
  {title} {Animal-assisted therapy and loneliness in nursing homes: use of
  robotic versus living dogs},}\ }\href@noop {} {\bibfield  {journal} {\bibinfo
   {journal} {Journal of the American Medical Directors Association}\ }\textbf
  {\bibinfo {volume} {9}},\ \bibinfo {pages} {173--177} (\bibinfo {year}
  {2008})}\BibitemShut {NoStop}%
\bibitem [{\citenamefont {Takayanagi}\ \emph {et~al.}(2014)\citenamefont
  {Takayanagi}, \citenamefont {Kirita},\ and\ \citenamefont
  {Shibata}}]{takayanagi2014comparison}%
  \BibitemOpen
  \bibfield  {author} {\bibinfo {author} {\bibfnamefont {Kazue}\ \bibnamefont
  {Takayanagi}}, \bibinfo {author} {\bibfnamefont {Takahiro}\ \bibnamefont
  {Kirita}}, \ and\ \bibinfo {author} {\bibfnamefont {Takanori}\ \bibnamefont
  {Shibata}},\ }\bibfield  {title} {\enquote {\bibinfo {title} {Comparison of
  verbal and emotional responses of elderly people with mild/moderate dementia
  and those with severe dementia in responses to seal robot, paro},}\
  }\href@noop {} {\bibfield  {journal} {\bibinfo  {journal} {Frontiers in Aging
  Neuroscience}\ }\textbf {\bibinfo {volume} {6}},\ \bibinfo {pages} {257}
  (\bibinfo {year} {2014})}\BibitemShut {NoStop}%
\bibitem [{\citenamefont {Wada}\ and\ \citenamefont
  {Shibata}(2007{\natexlab{a}})}]{wada2007social}%
  \BibitemOpen
  \bibfield  {author} {\bibinfo {author} {\bibfnamefont {Kazuyoshi}\
  \bibnamefont {Wada}}\ and\ \bibinfo {author} {\bibfnamefont {Takanori}\
  \bibnamefont {Shibata}},\ }\bibfield  {title} {\enquote {\bibinfo {title}
  {Social effects of robot therapy in a care house-change of social network of
  the residents for two months},}\ }in\ \href@noop {} {\emph {\bibinfo
  {booktitle} {Proceedings 2007 IEEE International Conference on Robotics and
  Automation}}}\ (\bibinfo {organization} {IEEE},\ \bibinfo {year} {2007})\
  pp.\ \bibinfo {pages} {1250--1255}\BibitemShut {NoStop}%
\bibitem [{\citenamefont {Arkin}\ \emph {et~al.}(2014)\citenamefont {Arkin},
  \citenamefont {Scheutz},\ and\ \citenamefont
  {Tickle-Degnen}}]{arkin2014preserving}%
  \BibitemOpen
  \bibfield  {author} {\bibinfo {author} {\bibfnamefont {Ronald~C}\
  \bibnamefont {Arkin}}, \bibinfo {author} {\bibfnamefont {Matthias}\
  \bibnamefont {Scheutz}}, \ and\ \bibinfo {author} {\bibfnamefont {Linda}\
  \bibnamefont {Tickle-Degnen}},\ }\bibfield  {title} {\enquote {\bibinfo
  {title} {Preserving dignity in patient caregiver relationships using moral
  emotions and robots},}\ }in\ \href@noop {} {\emph {\bibinfo {booktitle} {2014
  IEEE International Symposium on Ethics in Science, Technology and
  Engineering}}}\ (\bibinfo {organization} {IEEE},\ \bibinfo {year} {2014})\
  pp.\ \bibinfo {pages} {1--5}\BibitemShut {NoStop}%
\bibitem [{\citenamefont {Williams}\ \emph {et~al.}(2019)\citenamefont
  {Williams}, \citenamefont {Williams}, \citenamefont {Moore},\ and\
  \citenamefont {McFarlane}}]{williams2019aida}%
  \BibitemOpen
  \bibfield  {author} {\bibinfo {author} {\bibfnamefont {Andrew~B}\
  \bibnamefont {Williams}}, \bibinfo {author} {\bibfnamefont {Rosa~M}\
  \bibnamefont {Williams}}, \bibinfo {author} {\bibfnamefont {Ronald~E}\
  \bibnamefont {Moore}}, \ and\ \bibinfo {author} {\bibfnamefont {Matthias}\
  \bibnamefont {McFarlane}},\ }\bibfield  {title} {\enquote {\bibinfo {title}
  {Aida: A social co-robot to uplift workers with intellectual and
  developmental disabilities},}\ }in\ \href@noop {} {\emph {\bibinfo
  {booktitle} {2019 14th ACM/IEEE International Conference on Human-Robot
  Interaction (HRI)}}}\ (\bibinfo {organization} {IEEE},\ \bibinfo {year}
  {2019})\ pp.\ \bibinfo {pages} {584--585}\BibitemShut {NoStop}%
\bibitem [{\citenamefont {Moharana}\ \emph {et~al.}(2019)\citenamefont
  {Moharana}, \citenamefont {Panduro}, \citenamefont {Lee},\ and\ \citenamefont
  {Riek}}]{moharana2019robots}%
  \BibitemOpen
  \bibfield  {author} {\bibinfo {author} {\bibfnamefont {Sanika}\ \bibnamefont
  {Moharana}}, \bibinfo {author} {\bibfnamefont {Alejandro~E}\ \bibnamefont
  {Panduro}}, \bibinfo {author} {\bibfnamefont {Hee~Rin}\ \bibnamefont {Lee}},
  \ and\ \bibinfo {author} {\bibfnamefont {Laurel~D}\ \bibnamefont {Riek}},\
  }\bibfield  {title} {\enquote {\bibinfo {title} {Robots for joy, robots for
  sorrow: Community based robot design for dementia caregivers},}\ }in\
  \href@noop {} {\emph {\bibinfo {booktitle} {2019 14th ACM/IEEE International
  Conference on Human-Robot Interaction (HRI)}}}\ (\bibinfo {organization}
  {IEEE},\ \bibinfo {year} {2019})\ pp.\ \bibinfo {pages}
  {458--467}\BibitemShut {NoStop}%
\bibitem [{\citenamefont {Robins}\ \emph {et~al.}(2007)\citenamefont {Robins},
  \citenamefont {Otero}, \citenamefont {Ferrari},\ and\ \citenamefont
  {Dautenhahn}}]{robins2007eliciting}%
  \BibitemOpen
  \bibfield  {author} {\bibinfo {author} {\bibfnamefont {Ben}\ \bibnamefont
  {Robins}}, \bibinfo {author} {\bibfnamefont {Nuno}\ \bibnamefont {Otero}},
  \bibinfo {author} {\bibfnamefont {Ester}\ \bibnamefont {Ferrari}}, \ and\
  \bibinfo {author} {\bibfnamefont {Kerstin}\ \bibnamefont {Dautenhahn}},\
  }\bibfield  {title} {\enquote {\bibinfo {title} {Eliciting requirements for a
  robotic toy for children with autism-results from user panels},}\ }in\
  \href@noop {} {\emph {\bibinfo {booktitle} {RO-MAN 2007-The 16th IEEE
  International Symposium on Robot and Human Interactive Communication}}}\
  (\bibinfo {organization} {IEEE},\ \bibinfo {year} {2007})\ pp.\ \bibinfo
  {pages} {101--106}\BibitemShut {NoStop}%
\bibitem [{\citenamefont {Robins}\ \emph {et~al.}(2008)\citenamefont {Robins},
  \citenamefont {Ferrari},\ and\ \citenamefont
  {Dautenhahn}}]{robins2008developing}%
  \BibitemOpen
  \bibfield  {author} {\bibinfo {author} {\bibfnamefont {Ben}\ \bibnamefont
  {Robins}}, \bibinfo {author} {\bibfnamefont {Ester}\ \bibnamefont {Ferrari}},
  \ and\ \bibinfo {author} {\bibfnamefont {Kerstin}\ \bibnamefont
  {Dautenhahn}},\ }\bibfield  {title} {\enquote {\bibinfo {title} {Developing
  scenarios for robot assisted play},}\ }in\ \href@noop {} {\emph {\bibinfo
  {booktitle} {RO-MAN 2008-The 17th IEEE International Symposium on Robot and
  Human Interactive Communication}}}\ (\bibinfo {organization} {IEEE},\
  \bibinfo {year} {2008})\ pp.\ \bibinfo {pages} {180--186}\BibitemShut
  {NoStop}%
\bibitem [{\citenamefont {Short}\ and\ \citenamefont
  {Mataric}(2017)}]{short2017robot}%
  \BibitemOpen
  \bibfield  {author} {\bibinfo {author} {\bibfnamefont {Elaine}\ \bibnamefont
  {Short}}\ and\ \bibinfo {author} {\bibfnamefont {Maja~J}\ \bibnamefont
  {Mataric}},\ }\bibfield  {title} {\enquote {\bibinfo {title} {Robot
  moderation of a collaborative game: Towards socially assistive robotics in
  group interactions},}\ }in\ \href@noop {} {\emph {\bibinfo {booktitle} {2017
  26th IEEE International Symposium on Robot and Human Interactive
  Communication (RO-MAN)}}}\ (\bibinfo {organization} {IEEE},\ \bibinfo {year}
  {2017})\ pp.\ \bibinfo {pages} {385--390}\BibitemShut {NoStop}%
\bibitem [{\citenamefont {Matsuyama}\ \emph {et~al.}(2015)\citenamefont
  {Matsuyama}, \citenamefont {Akiba}, \citenamefont {Fujie},\ and\
  \citenamefont {Kobayashi}}]{matsuyama2015four}%
  \BibitemOpen
  \bibfield  {author} {\bibinfo {author} {\bibfnamefont {Yoichi}\ \bibnamefont
  {Matsuyama}}, \bibinfo {author} {\bibfnamefont {Iwao}\ \bibnamefont {Akiba}},
  \bibinfo {author} {\bibfnamefont {Shinya}\ \bibnamefont {Fujie}}, \ and\
  \bibinfo {author} {\bibfnamefont {Tetsunori}\ \bibnamefont {Kobayashi}},\
  }\bibfield  {title} {\enquote {\bibinfo {title} {Four-participant group
  conversation: A facilitation robot controlling engagement density as the
  fourth participant},}\ }\href@noop {} {\bibfield  {journal} {\bibinfo
  {journal} {Computer Speech \& Language}\ }\textbf {\bibinfo {volume} {33}},\
  \bibinfo {pages} {1--24} (\bibinfo {year} {2015})}\BibitemShut {NoStop}%
\bibitem [{\citenamefont {Chandra}\ \emph {et~al.}(2015)\citenamefont
  {Chandra}, \citenamefont {Alves-Oliveira}, \citenamefont {Lemaignan},
  \citenamefont {Sequeira}, \citenamefont {Paiva},\ and\ \citenamefont
  {Dillenbourg}}]{chandra2015can}%
  \BibitemOpen
  \bibfield  {author} {\bibinfo {author} {\bibfnamefont {Shruti}\ \bibnamefont
  {Chandra}}, \bibinfo {author} {\bibfnamefont {Patricia}\ \bibnamefont
  {Alves-Oliveira}}, \bibinfo {author} {\bibfnamefont {S{\'e}verin}\
  \bibnamefont {Lemaignan}}, \bibinfo {author} {\bibfnamefont {Pedro}\
  \bibnamefont {Sequeira}}, \bibinfo {author} {\bibfnamefont {Ana}\
  \bibnamefont {Paiva}}, \ and\ \bibinfo {author} {\bibfnamefont {Pierre}\
  \bibnamefont {Dillenbourg}},\ }\bibfield  {title} {\enquote {\bibinfo {title}
  {Can a child feel responsible for another in the presence of a robot in a
  collaborative learning activity?}}\ }in\ \href@noop {} {\emph {\bibinfo
  {booktitle} {2015 24th IEEE international symposium on robot and human
  interactive communication (RO-MAN)}}}\ (\bibinfo {organization} {IEEE},\
  \bibinfo {year} {2015})\ pp.\ \bibinfo {pages} {167--172}\BibitemShut
  {NoStop}%
\bibitem [{\citenamefont {Shen}\ \emph {et~al.}(2018)\citenamefont {Shen},
  \citenamefont {Slovak},\ and\ \citenamefont {Jung}}]{shen2018stop}%
  \BibitemOpen
  \bibfield  {author} {\bibinfo {author} {\bibfnamefont {Solace}\ \bibnamefont
  {Shen}}, \bibinfo {author} {\bibfnamefont {Petr}\ \bibnamefont {Slovak}}, \
  and\ \bibinfo {author} {\bibfnamefont {Malte~F}\ \bibnamefont {Jung}},\
  }\bibfield  {title} {\enquote {\bibinfo {title} {Stop. i see a conflict
  happening.: A robot mediator for young children's interpersonal conflict
  resolution},}\ }in\ \href@noop {} {\emph {\bibinfo {booktitle} {Proceedings
  of the 2018 ACM/IEEE International Conference on Human-Robot Interaction}}}\
  (\bibinfo {organization} {ACM},\ \bibinfo {year} {2018})\ pp.\ \bibinfo
  {pages} {69--77}\BibitemShut {NoStop}%
\bibitem [{\citenamefont {Mutlu}\ \emph {et~al.}(2009)\citenamefont {Mutlu},
  \citenamefont {Shiwa}, \citenamefont {Kanda}, \citenamefont {Ishiguro},\ and\
  \citenamefont {Hagita}}]{mutlu2009footing}%
  \BibitemOpen
  \bibfield  {author} {\bibinfo {author} {\bibfnamefont {Bilge}\ \bibnamefont
  {Mutlu}}, \bibinfo {author} {\bibfnamefont {Toshiyuki}\ \bibnamefont
  {Shiwa}}, \bibinfo {author} {\bibfnamefont {Takayuki}\ \bibnamefont {Kanda}},
  \bibinfo {author} {\bibfnamefont {Hiroshi}\ \bibnamefont {Ishiguro}}, \ and\
  \bibinfo {author} {\bibfnamefont {Norihiro}\ \bibnamefont {Hagita}},\
  }\bibfield  {title} {\enquote {\bibinfo {title} {Footing in human-robot
  conversations: how robots might shape participant roles using gaze cues},}\
  }in\ \href@noop {} {\emph {\bibinfo {booktitle} {Proceedings of the 4th
  ACM/IEEE international conference on Human robot interaction}}}\ (\bibinfo
  {organization} {ACM},\ \bibinfo {year} {2009})\ pp.\ \bibinfo {pages}
  {61--68}\BibitemShut {NoStop}%
\bibitem [{\citenamefont {Short}\ \emph {et~al.}(2016)\citenamefont {Short},
  \citenamefont {Sittig-Boyd},\ and\ \citenamefont
  {Mataric}}]{short2016modeling}%
  \BibitemOpen
  \bibfield  {author} {\bibinfo {author} {\bibfnamefont {Elaine}\ \bibnamefont
  {Short}}, \bibinfo {author} {\bibfnamefont {Katherine}\ \bibnamefont
  {Sittig-Boyd}}, \ and\ \bibinfo {author} {\bibfnamefont {Maja~J}\
  \bibnamefont {Mataric}},\ }\bibfield  {title} {\enquote {\bibinfo {title}
  {Modeling moderation for multi-party socially assistive robotics},}\ }in\
  \href@noop {} {\emph {\bibinfo {booktitle} {IEEE Int. Symp. Robot Hum.
  Interact. Commun.(RO-MAN 2016). New York, NY: IEEE}}}\ (\bibinfo {year}
  {2016})\BibitemShut {NoStop}%
\bibitem [{\citenamefont {Giannopulu}\ and\ \citenamefont
  {Pradel}(2012)}]{giannopulu2012child}%
  \BibitemOpen
  \bibfield  {author} {\bibinfo {author} {\bibfnamefont {Irini}\ \bibnamefont
  {Giannopulu}}\ and\ \bibinfo {author} {\bibfnamefont {Gilbert}\ \bibnamefont
  {Pradel}},\ }\bibfield  {title} {\enquote {\bibinfo {title} {From child-robot
  interaction to child-robot-therapist interaction: A case study in autism},}\
  }\href@noop {} {\bibfield  {journal} {\bibinfo  {journal} {Applied Bionics
  and Biomechanics}\ }\textbf {\bibinfo {volume} {9}},\ \bibinfo {pages}
  {173--179} (\bibinfo {year} {2012})}\BibitemShut {NoStop}%
\bibitem [{\citenamefont {Wood}\ \emph {et~al.}(2015)\citenamefont {Wood},
  \citenamefont {Sharkey}, \citenamefont {Mountain},\ and\ \citenamefont
  {Millings}}]{wood2015paro}%
  \BibitemOpen
  \bibfield  {author} {\bibinfo {author} {\bibfnamefont {Natalie}\ \bibnamefont
  {Wood}}, \bibinfo {author} {\bibfnamefont {Amanda}\ \bibnamefont {Sharkey}},
  \bibinfo {author} {\bibfnamefont {Gail}\ \bibnamefont {Mountain}}, \ and\
  \bibinfo {author} {\bibfnamefont {Abigail}\ \bibnamefont {Millings}},\
  }\bibfield  {title} {\enquote {\bibinfo {title} {The paro robot seal as a
  social mediator for healthy users},}\ }in\ \href@noop {} {\emph {\bibinfo
  {booktitle} {Proceedings of AISB Convention 2015}}}\ (\bibinfo {organization}
  {University of Kent},\ \bibinfo {year} {2015})\BibitemShut {NoStop}%
\bibitem [{\citenamefont {Kidd}\ \emph {et~al.}(2006)\citenamefont {Kidd},
  \citenamefont {Taggart},\ and\ \citenamefont {Turkle}}]{kidd2006sociable}%
  \BibitemOpen
  \bibfield  {author} {\bibinfo {author} {\bibfnamefont {Cory~D}\ \bibnamefont
  {Kidd}}, \bibinfo {author} {\bibfnamefont {Will}\ \bibnamefont {Taggart}}, \
  and\ \bibinfo {author} {\bibfnamefont {Sherry}\ \bibnamefont {Turkle}},\
  }\bibfield  {title} {\enquote {\bibinfo {title} {A sociable robot to
  encourage social interaction among the elderly},}\ }in\ \href@noop {} {\emph
  {\bibinfo {booktitle} {Proceedings 2006 IEEE International Conference on
  Robotics and Automation, 2006. ICRA 2006.}}}\ (\bibinfo {organization}
  {IEEE},\ \bibinfo {year} {2006})\ pp.\ \bibinfo {pages}
  {3972--3976}\BibitemShut {NoStop}%
\bibitem [{\citenamefont {Robins}\ \emph {et~al.}(2009)\citenamefont {Robins},
  \citenamefont {Dautenhahn},\ and\ \citenamefont
  {Dickerson}}]{robins2009isolation}%
  \BibitemOpen
  \bibfield  {author} {\bibinfo {author} {\bibfnamefont {Ben}\ \bibnamefont
  {Robins}}, \bibinfo {author} {\bibfnamefont {Kerstin}\ \bibnamefont
  {Dautenhahn}}, \ and\ \bibinfo {author} {\bibfnamefont {Paul}\ \bibnamefont
  {Dickerson}},\ }\bibfield  {title} {\enquote {\bibinfo {title} {From
  isolation to communication: a case study evaluation of robot assisted play
  for children with autism with a minimally expressive humanoid robot},}\ }in\
  \href@noop {} {\emph {\bibinfo {booktitle} {2009 Second International
  Conferences on Advances in Computer-Human Interactions}}}\ (\bibinfo
  {organization} {IEEE},\ \bibinfo {year} {2009})\ pp.\ \bibinfo {pages}
  {205--211}\BibitemShut {NoStop}%
\bibitem [{\citenamefont {Utami}\ and\ \citenamefont
  {Bickmore}(2019)}]{utami2019collaborative}%
  \BibitemOpen
  \bibfield  {author} {\bibinfo {author} {\bibfnamefont {Dina}\ \bibnamefont
  {Utami}}\ and\ \bibinfo {author} {\bibfnamefont {Timothy}\ \bibnamefont
  {Bickmore}},\ }\bibfield  {title} {\enquote {\bibinfo {title} {Collaborative
  user responses in multiparty interaction with a couples counselor robot},}\
  }in\ \href@noop {} {\emph {\bibinfo {booktitle} {2019 14th ACM/IEEE
  International Conference on Human-Robot Interaction (HRI)}}}\ (\bibinfo
  {organization} {IEEE},\ \bibinfo {year} {2019})\ pp.\ \bibinfo {pages}
  {294--303}\BibitemShut {NoStop}%
\bibitem [{\citenamefont {Short}\ \emph {et~al.}(2017)\citenamefont {Short},
  \citenamefont {Swift-Spong}, \citenamefont {Shim}, \citenamefont
  {Wisniewski}, \citenamefont {Zak}, \citenamefont {Wu}, \citenamefont
  {Zelinski},\ and\ \citenamefont {Matari{\'c}}}]{short2017understanding}%
  \BibitemOpen
  \bibfield  {author} {\bibinfo {author} {\bibfnamefont {Elaine~Schaertl}\
  \bibnamefont {Short}}, \bibinfo {author} {\bibfnamefont {Katelyn}\
  \bibnamefont {Swift-Spong}}, \bibinfo {author} {\bibfnamefont {Hyunju}\
  \bibnamefont {Shim}}, \bibinfo {author} {\bibfnamefont {Kristi~M}\
  \bibnamefont {Wisniewski}}, \bibinfo {author} {\bibfnamefont {Deanah~Kim}\
  \bibnamefont {Zak}}, \bibinfo {author} {\bibfnamefont {Shinyi}\ \bibnamefont
  {Wu}}, \bibinfo {author} {\bibfnamefont {Elizabeth}\ \bibnamefont
  {Zelinski}}, \ and\ \bibinfo {author} {\bibfnamefont {Maja~J}\ \bibnamefont
  {Matari{\'c}}},\ }\bibfield  {title} {\enquote {\bibinfo {title}
  {Understanding social interactions with socially assistive robotics in
  intergenerational family groups},}\ }in\ \href@noop {} {\emph {\bibinfo
  {booktitle} {2017 26th IEEE International Symposium on Robot and Human
  Interactive Communication (RO-MAN)}}}\ (\bibinfo {organization} {IEEE},\
  \bibinfo {year} {2017})\ pp.\ \bibinfo {pages} {236--241}\BibitemShut
  {NoStop}%
\bibitem [{\citenamefont {Wada}\ and\ \citenamefont
  {Shibata}(2006)}]{wada2006robot}%
  \BibitemOpen
  \bibfield  {author} {\bibinfo {author} {\bibfnamefont {Kazuyoshi}\
  \bibnamefont {Wada}}\ and\ \bibinfo {author} {\bibfnamefont {Takanori}\
  \bibnamefont {Shibata}},\ }\bibfield  {title} {\enquote {\bibinfo {title}
  {Robot therapy in a care house-its sociopsychological and physiological
  effects on the residents},}\ }in\ \href@noop {} {\emph {\bibinfo {booktitle}
  {Proceedings 2006 IEEE International Conference on Robotics and Automation,
  2006. ICRA 2006.}}}\ (\bibinfo {organization} {IEEE},\ \bibinfo {year}
  {2006})\ pp.\ \bibinfo {pages} {3966--3971}\BibitemShut {NoStop}%
\bibitem [{\citenamefont {Tickle-Degnen}\ \emph {et~al.}(2011)\citenamefont
  {Tickle-Degnen}, \citenamefont {Zebrowitz},\ and\ \citenamefont
  {Ma}}]{tickle2011culture}%
  \BibitemOpen
  \bibfield  {author} {\bibinfo {author} {\bibfnamefont {Linda}\ \bibnamefont
  {Tickle-Degnen}}, \bibinfo {author} {\bibfnamefont {Leslie~A}\ \bibnamefont
  {Zebrowitz}}, \ and\ \bibinfo {author} {\bibfnamefont {Hui-ing}\ \bibnamefont
  {Ma}},\ }\bibfield  {title} {\enquote {\bibinfo {title} {Culture, gender and
  health care stigma: Practitioners? response to facial masking experienced by
  people with parkinson?s disease},}\ }\href@noop {} {\bibfield  {journal}
  {\bibinfo  {journal} {Social Science \& Medicine}\ }\textbf {\bibinfo
  {volume} {73}},\ \bibinfo {pages} {95--102} (\bibinfo {year}
  {2011})}\BibitemShut {NoStop}%
\bibitem [{\citenamefont {Hemmesch}\ \emph {et~al.}(2009)\citenamefont
  {Hemmesch}, \citenamefont {Tickle-Degnen},\ and\ \citenamefont
  {Zebrowitz}}]{hemmesch2009influence}%
  \BibitemOpen
  \bibfield  {author} {\bibinfo {author} {\bibfnamefont {Amanda~R}\
  \bibnamefont {Hemmesch}}, \bibinfo {author} {\bibfnamefont {Linda}\
  \bibnamefont {Tickle-Degnen}}, \ and\ \bibinfo {author} {\bibfnamefont
  {Leslie~A}\ \bibnamefont {Zebrowitz}},\ }\bibfield  {title} {\enquote
  {\bibinfo {title} {The influence of facial masking and sex on older adults?
  impressions of individuals with parkinson?s disease.}}\ }\href@noop {}
  {\bibfield  {journal} {\bibinfo  {journal} {Psychology and aging}\ }\textbf
  {\bibinfo {volume} {24}},\ \bibinfo {pages} {542} (\bibinfo {year}
  {2009})}\BibitemShut {NoStop}%
\bibitem [{\citenamefont {Harms}\ \emph {et~al.}(2010)\citenamefont {Harms},
  \citenamefont {Martin},\ and\ \citenamefont {Wallace}}]{harms2010facial}%
  \BibitemOpen
  \bibfield  {author} {\bibinfo {author} {\bibfnamefont {Madeline~B}\
  \bibnamefont {Harms}}, \bibinfo {author} {\bibfnamefont {Alex}\ \bibnamefont
  {Martin}}, \ and\ \bibinfo {author} {\bibfnamefont {Gregory~L}\ \bibnamefont
  {Wallace}},\ }\bibfield  {title} {\enquote {\bibinfo {title} {Facial emotion
  recognition in autism spectrum disorders: a review of behavioral and
  neuroimaging studies},}\ }\href@noop {} {\bibfield  {journal} {\bibinfo
  {journal} {Neuropsychology review}\ }\textbf {\bibinfo {volume} {20}},\
  \bibinfo {pages} {290--322} (\bibinfo {year} {2010})}\BibitemShut {NoStop}%
\bibitem [{\citenamefont {Karlawish}\ \emph {et~al.}(2002)\citenamefont
  {Karlawish}, \citenamefont {Casarett}, \citenamefont {Propert}, \citenamefont
  {James}, \citenamefont {Bioethics},\ and\ \citenamefont
  {Clark}}]{karlawish2002relationship}%
  \BibitemOpen
  \bibfield  {author} {\bibinfo {author} {\bibfnamefont {Jason~HT}\
  \bibnamefont {Karlawish}}, \bibinfo {author} {\bibfnamefont {David}\
  \bibnamefont {Casarett}}, \bibinfo {author} {\bibfnamefont {Kathleen~Joy}\
  \bibnamefont {Propert}}, \bibinfo {author} {\bibfnamefont {Bryan~D}\
  \bibnamefont {James}}, \bibinfo {author} {\bibfnamefont {M}~\bibnamefont
  {Bioethics}}, \ and\ \bibinfo {author} {\bibfnamefont {Christopher~M}\
  \bibnamefont {Clark}},\ }\bibfield  {title} {\enquote {\bibinfo {title}
  {Relationship between alzheimer's disease severity and patient participation
  in decisions about their medical care},}\ }\href@noop {} {\bibfield
  {journal} {\bibinfo  {journal} {Journal of geriatric psychiatry and
  neurology}\ }\textbf {\bibinfo {volume} {15}},\ \bibinfo {pages} {68--72}
  (\bibinfo {year} {2002})}\BibitemShut {NoStop}%
\bibitem [{\citenamefont {Sabat}(2005)}]{sabat2005capacity}%
  \BibitemOpen
  \bibfield  {author} {\bibinfo {author} {\bibfnamefont {Steven~R}\
  \bibnamefont {Sabat}},\ }\bibfield  {title} {\enquote {\bibinfo {title}
  {Capacity for decision-making in alzheimer's disease: Selfhood, positioning
  and semiotic people},}\ }\href@noop {} {\bibfield  {journal} {\bibinfo
  {journal} {Australian \& New Zealand Journal of Psychiatry}\ }\textbf
  {\bibinfo {volume} {39}},\ \bibinfo {pages} {1030--1035} (\bibinfo {year}
  {2005})}\BibitemShut {NoStop}%
\bibitem [{\citenamefont {Preston}\ \emph {et~al.}(2013)\citenamefont
  {Preston}, \citenamefont {Hofelich},\ and\ \citenamefont
  {Stansfield}}]{preston2013ethology}%
  \BibitemOpen
  \bibfield  {author} {\bibinfo {author} {\bibfnamefont {Stephanie~D}\
  \bibnamefont {Preston}}, \bibinfo {author} {\bibfnamefont {Alicia~J}\
  \bibnamefont {Hofelich}}, \ and\ \bibinfo {author} {\bibfnamefont {Robert~B}\
  \bibnamefont {Stansfield}},\ }\bibfield  {title} {\enquote {\bibinfo {title}
  {The ethology of empathy: a taxonomy of real-world targets of need and their
  effect on observers},}\ }\href@noop {} {\bibfield  {journal} {\bibinfo
  {journal} {Frontiers in human neuroscience}\ }\textbf {\bibinfo {volume}
  {7}},\ \bibinfo {pages} {488} (\bibinfo {year} {2013})}\BibitemShut {NoStop}%
\bibitem [{\citenamefont {Escher}\ \emph {et~al.}(2004)\citenamefont {Escher},
  \citenamefont {Perneger},\ and\ \citenamefont
  {Chevrolet}}]{escher2004national}%
  \BibitemOpen
  \bibfield  {author} {\bibinfo {author} {\bibfnamefont {Monica}\ \bibnamefont
  {Escher}}, \bibinfo {author} {\bibfnamefont {Thomas~V}\ \bibnamefont
  {Perneger}}, \ and\ \bibinfo {author} {\bibfnamefont {Jean-Claude}\
  \bibnamefont {Chevrolet}},\ }\bibfield  {title} {\enquote {\bibinfo {title}
  {National questionnaire survey on what influences doctors' decisions about
  admission to intensive care},}\ }\href@noop {} {\bibfield  {journal}
  {\bibinfo  {journal} {Bmj}\ }\textbf {\bibinfo {volume} {329}},\ \bibinfo
  {pages} {425} (\bibinfo {year} {2004})}\BibitemShut {NoStop}%
\bibitem [{\citenamefont {Gerbert}(1984)}]{gerbert1984perceived}%
  \BibitemOpen
  \bibfield  {author} {\bibinfo {author} {\bibfnamefont {Barbara}\ \bibnamefont
  {Gerbert}},\ }\bibfield  {title} {\enquote {\bibinfo {title} {Perceived
  likeability and competence of simulated patients: influence on physicians'
  management plans},}\ }\href@noop {} {\bibfield  {journal} {\bibinfo
  {journal} {Social science \& medicine}\ }\textbf {\bibinfo {volume} {18}},\
  \bibinfo {pages} {1053--1059} (\bibinfo {year} {1984})}\BibitemShut {NoStop}%
\bibitem [{\citenamefont {Chita-Tegmark}\ \emph {et~al.}(2019)\citenamefont
  {Chita-Tegmark}, \citenamefont {Ackerman},\ and\ \citenamefont
  {Scheutz}}]{chita2019effects}%
  \BibitemOpen
  \bibfield  {author} {\bibinfo {author} {\bibfnamefont {Meia}\ \bibnamefont
  {Chita-Tegmark}}, \bibinfo {author} {\bibfnamefont {Janet~M}\ \bibnamefont
  {Ackerman}}, \ and\ \bibinfo {author} {\bibfnamefont {Matthias}\ \bibnamefont
  {Scheutz}},\ }\bibfield  {title} {\enquote {\bibinfo {title} {Effects of
  assistive robot behavior on impressions of patient psychological attributes:
  Vignette-based human-robot interaction study},}\ }\href@noop {} {\bibfield
  {journal} {\bibinfo  {journal} {Journal of medical Internet research}\
  }\textbf {\bibinfo {volume} {21}},\ \bibinfo {pages} {e13729} (\bibinfo
  {year} {2019})}\BibitemShut {NoStop}%
\bibitem [{\citenamefont {Haque}\ and\ \citenamefont
  {Waytz}(2012)}]{haque2012dehumanization}%
  \BibitemOpen
  \bibfield  {author} {\bibinfo {author} {\bibfnamefont {Omar~Sultan}\
  \bibnamefont {Haque}}\ and\ \bibinfo {author} {\bibfnamefont {Adam}\
  \bibnamefont {Waytz}},\ }\bibfield  {title} {\enquote {\bibinfo {title}
  {Dehumanization in medicine: Causes, solutions, and functions},}\ }\href@noop
  {} {\bibfield  {journal} {\bibinfo  {journal} {Perspectives on psychological
  science}\ }\textbf {\bibinfo {volume} {7}},\ \bibinfo {pages} {176--186}
  (\bibinfo {year} {2012})}\BibitemShut {NoStop}%
\bibitem [{\citenamefont {Tickle-Degnen}\ and\ \citenamefont
  {Lyons}(2004)}]{tickle2004practitioners}%
  \BibitemOpen
  \bibfield  {author} {\bibinfo {author} {\bibfnamefont {Linda}\ \bibnamefont
  {Tickle-Degnen}}\ and\ \bibinfo {author} {\bibfnamefont {Kathleen~Doyle}\
  \bibnamefont {Lyons}},\ }\bibfield  {title} {\enquote {\bibinfo {title}
  {Practitioners? impressions of patients with parkinson's disease: the social
  ecology of the expressive mask},}\ }\href@noop {} {\bibfield  {journal}
  {\bibinfo  {journal} {Social Science \& Medicine}\ }\textbf {\bibinfo
  {volume} {58}},\ \bibinfo {pages} {603--614} (\bibinfo {year}
  {2004})}\BibitemShut {NoStop}%
\bibitem [{\citenamefont {Lyons}\ \emph {et~al.}(2004)\citenamefont {Lyons},
  \citenamefont {Tickle-Degnen}, \citenamefont {Henry},\ and\ \citenamefont
  {Cohn}}]{lyons2004impressions}%
  \BibitemOpen
  \bibfield  {author} {\bibinfo {author} {\bibfnamefont {Kathleen~Doyle}\
  \bibnamefont {Lyons}}, \bibinfo {author} {\bibfnamefont {Linda}\ \bibnamefont
  {Tickle-Degnen}}, \bibinfo {author} {\bibfnamefont {Alexis}\ \bibnamefont
  {Henry}}, \ and\ \bibinfo {author} {\bibfnamefont {Ellen}\ \bibnamefont
  {Cohn}},\ }\bibfield  {title} {\enquote {\bibinfo {title} {Impressions of
  personality in parkinson's disease: can rehabilitation practitioners see
  beyond the symptoms?}}\ }\href@noop {} {\bibfield  {journal} {\bibinfo
  {journal} {Rehabilitation Psychology}\ }\textbf {\bibinfo {volume} {49}},\
  \bibinfo {pages} {328} (\bibinfo {year} {2004})}\BibitemShut {NoStop}%
\bibitem [{\citenamefont {Baio}(2014)}]{baio2014prevalence}%
  \BibitemOpen
  \bibfield  {author} {\bibinfo {author} {\bibfnamefont {Jon}\ \bibnamefont
  {Baio}},\ }\bibfield  {title} {\enquote {\bibinfo {title} {Prevalence of
  autism spectrum disorder among children aged 8 years-autism and developmental
  disabilities monitoring network, 11 sites, united states, 2010},}\
  }\href@noop {} {\  (\bibinfo {year} {2014})}\BibitemShut {NoStop}%
\bibitem [{\citenamefont {Kim}\ \emph {et~al.}(2013{\natexlab{b}})\citenamefont
  {Kim}, \citenamefont {Berkovits}, \citenamefont {Bernier}, \citenamefont
  {Leyzberg}, \citenamefont {Shic}, \citenamefont {Paul},\ and\ \citenamefont
  {Scassellati}}]{kim2013social}%
  \BibitemOpen
  \bibfield  {author} {\bibinfo {author} {\bibfnamefont {Elizabeth~S}\
  \bibnamefont {Kim}}, \bibinfo {author} {\bibfnamefont {Lauren~D}\
  \bibnamefont {Berkovits}}, \bibinfo {author} {\bibfnamefont {Emily~P}\
  \bibnamefont {Bernier}}, \bibinfo {author} {\bibfnamefont {Dan}\ \bibnamefont
  {Leyzberg}}, \bibinfo {author} {\bibfnamefont {Frederick}\ \bibnamefont
  {Shic}}, \bibinfo {author} {\bibfnamefont {Rhea}\ \bibnamefont {Paul}}, \
  and\ \bibinfo {author} {\bibfnamefont {Brian}\ \bibnamefont {Scassellati}},\
  }\bibfield  {title} {\enquote {\bibinfo {title} {Social robots as embedded
  reinforcers of social behavior in children with autism},}\ }\href@noop {}
  {\bibfield  {journal} {\bibinfo  {journal} {Journal of autism and
  developmental disorders}\ }\textbf {\bibinfo {volume} {43}},\ \bibinfo
  {pages} {1038--1049} (\bibinfo {year} {2013}{\natexlab{b}})}\BibitemShut
  {NoStop}%
\bibitem [{\citenamefont {Segrin}(2000)}]{segrin2000social}%
  \BibitemOpen
  \bibfield  {author} {\bibinfo {author} {\bibfnamefont {Chris}\ \bibnamefont
  {Segrin}},\ }\bibfield  {title} {\enquote {\bibinfo {title} {Social skills
  deficits associated with depression},}\ }\href@noop {} {\bibfield  {journal}
  {\bibinfo  {journal} {Clinical psychology review}\ }\textbf {\bibinfo
  {volume} {20}},\ \bibinfo {pages} {379--403} (\bibinfo {year}
  {2000})}\BibitemShut {NoStop}%
\bibitem [{\citenamefont {Arkin}\ and\ \citenamefont
  {Pettinati}(2014)}]{arkin2014moral}%
  \BibitemOpen
  \bibfield  {author} {\bibinfo {author} {\bibfnamefont {Ronald~C}\
  \bibnamefont {Arkin}}\ and\ \bibinfo {author} {\bibfnamefont {Michael~J}\
  \bibnamefont {Pettinati}},\ }\bibfield  {title} {\enquote {\bibinfo {title}
  {Moral emotions, robots, and their role in managing stigma in early stage
  parkinson?s disease caregiving},}\ }\href@noop {} {\  (\bibinfo {year}
  {2014})}\BibitemShut {NoStop}%
\bibitem [{\citenamefont {Haslam}\ and\ \citenamefont
  {Loughnan}(2014)}]{haslam2014dehumanization}%
  \BibitemOpen
  \bibfield  {author} {\bibinfo {author} {\bibfnamefont {Nick}\ \bibnamefont
  {Haslam}}\ and\ \bibinfo {author} {\bibfnamefont {Steve}\ \bibnamefont
  {Loughnan}},\ }\bibfield  {title} {\enquote {\bibinfo {title} {Dehumanization
  and infrahumanization},}\ }\href@noop {} {\bibfield  {journal} {\bibinfo
  {journal} {Annual review of psychology}\ }\textbf {\bibinfo {volume} {65}},\
  \bibinfo {pages} {399--423} (\bibinfo {year} {2014})}\BibitemShut {NoStop}%
\bibitem [{\citenamefont {Decety}\ \emph {et~al.}(2010)\citenamefont {Decety},
  \citenamefont {Yang},\ and\ \citenamefont {Cheng}}]{decety2010physicians}%
  \BibitemOpen
  \bibfield  {author} {\bibinfo {author} {\bibfnamefont {Jean}\ \bibnamefont
  {Decety}}, \bibinfo {author} {\bibfnamefont {Chia-Yan}\ \bibnamefont {Yang}},
  \ and\ \bibinfo {author} {\bibfnamefont {Yawei}\ \bibnamefont {Cheng}},\
  }\bibfield  {title} {\enquote {\bibinfo {title} {Physicians down-regulate
  their pain empathy response: an event-related brain potential study},}\
  }\href@noop {} {\bibfield  {journal} {\bibinfo  {journal} {Neuroimage}\
  }\textbf {\bibinfo {volume} {50}},\ \bibinfo {pages} {1676--1682} (\bibinfo
  {year} {2010})}\BibitemShut {NoStop}%
\bibitem [{\citenamefont {Cheng}\ \emph {et~al.}(2007)\citenamefont {Cheng},
  \citenamefont {Lin}, \citenamefont {Liu}, \citenamefont {Hsu}, \citenamefont
  {Lim}, \citenamefont {Hung},\ and\ \citenamefont
  {Decety}}]{cheng2007expertise}%
  \BibitemOpen
  \bibfield  {author} {\bibinfo {author} {\bibfnamefont {Yawei}\ \bibnamefont
  {Cheng}}, \bibinfo {author} {\bibfnamefont {Ching-Po}\ \bibnamefont {Lin}},
  \bibinfo {author} {\bibfnamefont {Ho-Ling}\ \bibnamefont {Liu}}, \bibinfo
  {author} {\bibfnamefont {Yuan-Yu}\ \bibnamefont {Hsu}}, \bibinfo {author}
  {\bibfnamefont {Kun-Eng}\ \bibnamefont {Lim}}, \bibinfo {author}
  {\bibfnamefont {Daisy}\ \bibnamefont {Hung}}, \ and\ \bibinfo {author}
  {\bibfnamefont {Jean}\ \bibnamefont {Decety}},\ }\bibfield  {title} {\enquote
  {\bibinfo {title} {Expertise modulates the perception of pain in others},}\
  }\href@noop {} {\bibfield  {journal} {\bibinfo  {journal} {Current Biology}\
  }\textbf {\bibinfo {volume} {17}},\ \bibinfo {pages} {1708--1713} (\bibinfo
  {year} {2007})}\BibitemShut {NoStop}%
\bibitem [{\citenamefont {Rathert}\ \emph {et~al.}(2013)\citenamefont
  {Rathert}, \citenamefont {Wyrwich},\ and\ \citenamefont
  {Boren}}]{rathert2013patient}%
  \BibitemOpen
  \bibfield  {author} {\bibinfo {author} {\bibfnamefont {Cheryl}\ \bibnamefont
  {Rathert}}, \bibinfo {author} {\bibfnamefont {Mary~D}\ \bibnamefont
  {Wyrwich}}, \ and\ \bibinfo {author} {\bibfnamefont {Suzanne~Austin}\
  \bibnamefont {Boren}},\ }\bibfield  {title} {\enquote {\bibinfo {title}
  {Patient-centered care and outcomes: a systematic review of the
  literature},}\ }\href@noop {} {\bibfield  {journal} {\bibinfo  {journal}
  {Medical Care Research and Review}\ }\textbf {\bibinfo {volume} {70}},\
  \bibinfo {pages} {351--379} (\bibinfo {year} {2013})}\BibitemShut {NoStop}%
\bibitem [{\citenamefont {Stewart}(2001)}]{stewart2001towards}%
  \BibitemOpen
  \bibfield  {author} {\bibinfo {author} {\bibfnamefont {Moira}\ \bibnamefont
  {Stewart}},\ }\bibfield  {title} {\enquote {\bibinfo {title} {Towards a
  global definition of patient centred care},}\ }\href {\doibase
  10.1136/bmj.322.7284.444} {\bibfield  {journal} {\bibinfo  {journal} {BMJ}\
  }\textbf {\bibinfo {volume} {322}},\ \bibinfo {pages} {444--445} (\bibinfo
  {year} {2001})},\ \Eprint
  {http://arxiv.org/abs/https://www.bmj.com/content/322/7284/444.full.pdf}
  {https://www.bmj.com/content/322/7284/444.full.pdf} \BibitemShut {NoStop}%
\bibitem [{\citenamefont {Constand}\ \emph {et~al.}(2014)\citenamefont
  {Constand}, \citenamefont {MacDermid}, \citenamefont {Dal Bello-Haas},\ and\
  \citenamefont {Law}}]{constand2014scoping}%
  \BibitemOpen
  \bibfield  {author} {\bibinfo {author} {\bibfnamefont {Marissa~K}\
  \bibnamefont {Constand}}, \bibinfo {author} {\bibfnamefont {Joy~C}\
  \bibnamefont {MacDermid}}, \bibinfo {author} {\bibfnamefont {Vanina}\
  \bibnamefont {Dal Bello-Haas}}, \ and\ \bibinfo {author} {\bibfnamefont
  {Mary}\ \bibnamefont {Law}},\ }\bibfield  {title} {\enquote {\bibinfo {title}
  {Scoping review of patient-centered care approaches in healthcare},}\
  }\href@noop {} {\bibfield  {journal} {\bibinfo  {journal} {BMC health
  services research}\ }\textbf {\bibinfo {volume} {14}},\ \bibinfo {pages}
  {271} (\bibinfo {year} {2014})}\BibitemShut {NoStop}%
\bibitem [{\citenamefont {Haslam}(2007)}]{haslam2007humanising}%
  \BibitemOpen
  \bibfield  {author} {\bibinfo {author} {\bibfnamefont {Nick}\ \bibnamefont
  {Haslam}},\ }\bibfield  {title} {\enquote {\bibinfo {title} {Humanising
  medical practice: the role of empathy},}\ }\href@noop {} {\bibfield
  {journal} {\bibinfo  {journal} {Medical journal of Australia}\ }\textbf
  {\bibinfo {volume} {187}},\ \bibinfo {pages} {381--382} (\bibinfo {year}
  {2007})}\BibitemShut {NoStop}%
\bibitem [{\citenamefont {Riess}\ and\ \citenamefont
  {Kraft-Todd}(2014)}]{riess2014empathy}%
  \BibitemOpen
  \bibfield  {author} {\bibinfo {author} {\bibfnamefont {Helen}\ \bibnamefont
  {Riess}}\ and\ \bibinfo {author} {\bibfnamefont {Gordon}\ \bibnamefont
  {Kraft-Todd}},\ }\bibfield  {title} {\enquote {\bibinfo {title} {Empathy: a
  tool to enhance nonverbal communication between clinicians and their
  patients},}\ }\href@noop {} {\bibfield  {journal} {\bibinfo  {journal}
  {Academic Medicine}\ }\textbf {\bibinfo {volume} {89}},\ \bibinfo {pages}
  {1108--1112} (\bibinfo {year} {2014})}\BibitemShut {NoStop}%
\bibitem [{\citenamefont {Hoffman}\ \emph {et~al.}(2015)\citenamefont
  {Hoffman}, \citenamefont {Zuckerman}, \citenamefont {Hirschberger},
  \citenamefont {Luria},\ and\ \citenamefont
  {Shani~Sherman}}]{hoffman2015design}%
  \BibitemOpen
  \bibfield  {author} {\bibinfo {author} {\bibfnamefont {Guy}\ \bibnamefont
  {Hoffman}}, \bibinfo {author} {\bibfnamefont {Oren}\ \bibnamefont
  {Zuckerman}}, \bibinfo {author} {\bibfnamefont {Gilad}\ \bibnamefont
  {Hirschberger}}, \bibinfo {author} {\bibfnamefont {Michal}\ \bibnamefont
  {Luria}}, \ and\ \bibinfo {author} {\bibfnamefont {Tal}\ \bibnamefont
  {Shani~Sherman}},\ }\bibfield  {title} {\enquote {\bibinfo {title} {Design
  and evaluation of a peripheral robotic conversation companion},}\ }in\
  \href@noop {} {\emph {\bibinfo {booktitle} {Proceedings of the Tenth Annual
  ACM/IEEE International Conference on Human-Robot Interaction}}}\ (\bibinfo
  {organization} {ACM},\ \bibinfo {year} {2015})\ pp.\ \bibinfo {pages}
  {3--10}\BibitemShut {NoStop}%
\bibitem [{\citenamefont {Tahir}\ \emph {et~al.}(2014)\citenamefont {Tahir},
  \citenamefont {Rasheed}, \citenamefont {Dauwels},\ and\ \citenamefont
  {Dauwels}}]{tahir2014perception}%
  \BibitemOpen
  \bibfield  {author} {\bibinfo {author} {\bibfnamefont {Yasir}\ \bibnamefont
  {Tahir}}, \bibinfo {author} {\bibfnamefont {Umer}\ \bibnamefont {Rasheed}},
  \bibinfo {author} {\bibfnamefont {Shoko}\ \bibnamefont {Dauwels}}, \ and\
  \bibinfo {author} {\bibfnamefont {Justin}\ \bibnamefont {Dauwels}},\
  }\bibfield  {title} {\enquote {\bibinfo {title} {Perception of humanoid
  social mediator in two-person dialogs},}\ }in\ \href@noop {} {\emph {\bibinfo
  {booktitle} {Proceedings of the 2014 ACM/IEEE international conference on
  Human-robot interaction}}}\ (\bibinfo {organization} {ACM},\ \bibinfo {year}
  {2014})\ pp.\ \bibinfo {pages} {300--301}\BibitemShut {NoStop}%
\bibitem [{\citenamefont {Wilson}\ \emph
  {et~al.}(2016{\natexlab{b}})\citenamefont {Wilson}, \citenamefont {Arnold},\
  and\ \citenamefont {Scheutz}}]{wilson2016relational}%
  \BibitemOpen
  \bibfield  {author} {\bibinfo {author} {\bibfnamefont {Jason~R}\ \bibnamefont
  {Wilson}}, \bibinfo {author} {\bibfnamefont {Thomas}\ \bibnamefont {Arnold}},
  \ and\ \bibinfo {author} {\bibfnamefont {Matthias}\ \bibnamefont {Scheutz}},\
  }\bibfield  {title} {\enquote {\bibinfo {title} {Relational enhancement: A
  framework for evaluating and designing human-robot relationships},}\ }in\
  \href@noop {} {\emph {\bibinfo {booktitle} {Workshops at the Thirtieth AAAI
  Conference on Artificial Intelligence}}}\ (\bibinfo {year}
  {2016})\BibitemShut {NoStop}%
\bibitem [{\citenamefont {Shim}\ \emph {et~al.}(2017)\citenamefont {Shim},
  \citenamefont {Arkin},\ and\ \citenamefont
  {Pettinatti}}]{shim2017intervening}%
  \BibitemOpen
  \bibfield  {author} {\bibinfo {author} {\bibfnamefont {Jaeeun}\ \bibnamefont
  {Shim}}, \bibinfo {author} {\bibfnamefont {Ronald}\ \bibnamefont {Arkin}}, \
  and\ \bibinfo {author} {\bibfnamefont {Michael}\ \bibnamefont {Pettinatti}},\
  }\bibfield  {title} {\enquote {\bibinfo {title} {An intervening ethical
  governor for a robot mediator in patient-caregiver relationship:
  Implementation and evaluation},}\ }in\ \href@noop {} {\emph {\bibinfo
  {booktitle} {2017 IEEE International Conference on Robotics and Automation
  (ICRA)}}}\ (\bibinfo {organization} {IEEE},\ \bibinfo {year} {2017})\ pp.\
  \bibinfo {pages} {2936--2942}\BibitemShut {NoStop}%
\bibitem [{\citenamefont {Stoll}\ \emph {et~al.}(2018)\citenamefont {Stoll},
  \citenamefont {Jung},\ and\ \citenamefont {Fussell}}]{stoll2018keeping}%
  \BibitemOpen
  \bibfield  {author} {\bibinfo {author} {\bibfnamefont {Brett}\ \bibnamefont
  {Stoll}}, \bibinfo {author} {\bibfnamefont {Malte~F}\ \bibnamefont {Jung}}, \
  and\ \bibinfo {author} {\bibfnamefont {Susan~R}\ \bibnamefont {Fussell}},\
  }\bibfield  {title} {\enquote {\bibinfo {title} {Keeping it light:
  Perceptions of humor styles in robot-mediated conflict},}\ }in\ \href@noop {}
  {\emph {\bibinfo {booktitle} {Companion of the 2018 ACM/IEEE International
  Conference on Human-Robot Interaction}}}\ (\bibinfo {organization} {ACM},\
  \bibinfo {year} {2018})\ pp.\ \bibinfo {pages} {247--248}\BibitemShut
  {NoStop}%
\bibitem [{\citenamefont {Bippus}(2003)}]{bippus2003humor}%
  \BibitemOpen
  \bibfield  {author} {\bibinfo {author} {\bibfnamefont {Amy~M}\ \bibnamefont
  {Bippus}},\ }\bibfield  {title} {\enquote {\bibinfo {title} {Humor motives,
  qualities, and reactions in recalled conflict episodes},}\ }\href@noop {}
  {\bibfield  {journal} {\bibinfo  {journal} {Western Journal of Communication
  (includes Communication Reports)}\ }\textbf {\bibinfo {volume} {67}},\
  \bibinfo {pages} {413--426} (\bibinfo {year} {2003})}\BibitemShut {NoStop}%
\bibitem [{\citenamefont {Cacioppo}\ and\ \citenamefont
  {Hawkley}(2003)}]{cacioppo2003social}%
  \BibitemOpen
  \bibfield  {author} {\bibinfo {author} {\bibfnamefont {John~T}\ \bibnamefont
  {Cacioppo}}\ and\ \bibinfo {author} {\bibfnamefont {Louise~C}\ \bibnamefont
  {Hawkley}},\ }\bibfield  {title} {\enquote {\bibinfo {title} {Social
  isolation and health, with an emphasis on underlying mechanisms},}\
  }\href@noop {} {\bibfield  {journal} {\bibinfo  {journal} {Perspectives in
  biology and medicine}\ }\textbf {\bibinfo {volume} {46}},\ \bibinfo {pages}
  {S39--S52} (\bibinfo {year} {2003})}\BibitemShut {NoStop}%
\bibitem [{\citenamefont {Wada}\ and\ \citenamefont
  {Shibata}(2007{\natexlab{b}})}]{wada2007living}%
  \BibitemOpen
  \bibfield  {author} {\bibinfo {author} {\bibfnamefont {Kazuyoshi}\
  \bibnamefont {Wada}}\ and\ \bibinfo {author} {\bibfnamefont {Takanori}\
  \bibnamefont {Shibata}},\ }\bibfield  {title} {\enquote {\bibinfo {title}
  {Living with seal robots?its sociopsychological and physiological influences
  on the elderly at a care house},}\ }\href@noop {} {\bibfield  {journal}
  {\bibinfo  {journal} {IEEE transactions on robotics}\ }\textbf {\bibinfo
  {volume} {23}},\ \bibinfo {pages} {972--980} (\bibinfo {year}
  {2007}{\natexlab{b}})}\BibitemShut {NoStop}%
\bibitem [{\citenamefont {Robinson}\ \emph {et~al.}(2013)\citenamefont
  {Robinson}, \citenamefont {MacDonald}, \citenamefont {Kerse},\ and\
  \citenamefont {Broadbent}}]{robinson2013psychosocial}%
  \BibitemOpen
  \bibfield  {author} {\bibinfo {author} {\bibfnamefont {Hayley}\ \bibnamefont
  {Robinson}}, \bibinfo {author} {\bibfnamefont {Bruce}\ \bibnamefont
  {MacDonald}}, \bibinfo {author} {\bibfnamefont {Ngaire}\ \bibnamefont
  {Kerse}}, \ and\ \bibinfo {author} {\bibfnamefont {Elizabeth}\ \bibnamefont
  {Broadbent}},\ }\bibfield  {title} {\enquote {\bibinfo {title} {The
  psychosocial effects of a companion robot: a randomized controlled trial},}\
  }\href@noop {} {\bibfield  {journal} {\bibinfo  {journal} {Journal of the
  American Medical Directors Association}\ }\textbf {\bibinfo {volume} {14}},\
  \bibinfo {pages} {661--667} (\bibinfo {year} {2013})}\BibitemShut {NoStop}%
\bibitem [{\citenamefont {Joshi}\ and\ \citenamefont
  {{\v{S}}abanovi{\'c}}(2019)}]{joshi2019robots}%
  \BibitemOpen
  \bibfield  {author} {\bibinfo {author} {\bibfnamefont {Swapna}\ \bibnamefont
  {Joshi}}\ and\ \bibinfo {author} {\bibfnamefont {Selma}\ \bibnamefont
  {{\v{S}}abanovi{\'c}}},\ }\bibfield  {title} {\enquote {\bibinfo {title}
  {Robots for inter-generational interactions: Implications for nonfamilial
  community settings},}\ }in\ \href@noop {} {\emph {\bibinfo {booktitle} {2019
  14th ACM/IEEE International Conference on Human-Robot Interaction (HRI)}}}\
  (\bibinfo {organization} {IEEE},\ \bibinfo {year} {2019})\ pp.\ \bibinfo
  {pages} {478--486}\BibitemShut {NoStop}%
\bibitem [{\citenamefont {Werry}\ \emph {et~al.}(2001)\citenamefont {Werry},
  \citenamefont {Dautenhahn}, \citenamefont {Ogden},\ and\ \citenamefont
  {Harwin}}]{werry2001can}%
  \BibitemOpen
  \bibfield  {author} {\bibinfo {author} {\bibfnamefont {Iain}\ \bibnamefont
  {Werry}}, \bibinfo {author} {\bibfnamefont {Kerstin}\ \bibnamefont
  {Dautenhahn}}, \bibinfo {author} {\bibfnamefont {Bernard}\ \bibnamefont
  {Ogden}}, \ and\ \bibinfo {author} {\bibfnamefont {William}\ \bibnamefont
  {Harwin}},\ }\bibfield  {title} {\enquote {\bibinfo {title} {Can social
  interaction skills be taught by a social agent? the role of a robotic
  mediator in autism therapy},}\ }in\ \href@noop {} {\emph {\bibinfo
  {booktitle} {International Conference on Cognitive Technology}}}\ (\bibinfo
  {organization} {Springer},\ \bibinfo {year} {2001})\ pp.\ \bibinfo {pages}
  {57--74}\BibitemShut {NoStop}%
\bibitem [{\citenamefont {Miller}\ \emph {et~al.}(2006)\citenamefont {Miller},
  \citenamefont {Noble}, \citenamefont {Jones},\ and\ \citenamefont
  {Burn}}]{miller2006life}%
  \BibitemOpen
  \bibfield  {author} {\bibinfo {author} {\bibfnamefont {Nick}\ \bibnamefont
  {Miller}}, \bibinfo {author} {\bibfnamefont {Emma}\ \bibnamefont {Noble}},
  \bibinfo {author} {\bibfnamefont {Diana}\ \bibnamefont {Jones}}, \ and\
  \bibinfo {author} {\bibfnamefont {David}\ \bibnamefont {Burn}},\ }\bibfield
  {title} {\enquote {\bibinfo {title} {Life with communication changes in
  parkinson?s disease},}\ }\href@noop {} {\bibfield  {journal} {\bibinfo
  {journal} {Age and ageing}\ }\textbf {\bibinfo {volume} {35}},\ \bibinfo
  {pages} {235--239} (\bibinfo {year} {2006})}\BibitemShut {NoStop}%
\bibitem [{\citenamefont {McNamara}\ and\ \citenamefont
  {Durso}(2003)}]{mcnamara2003pragmatic}%
  \BibitemOpen
  \bibfield  {author} {\bibinfo {author} {\bibfnamefont {Patrick}\ \bibnamefont
  {McNamara}}\ and\ \bibinfo {author} {\bibfnamefont {Raymon}\ \bibnamefont
  {Durso}},\ }\bibfield  {title} {\enquote {\bibinfo {title} {Pragmatic
  communication skills in patients with parkinson?s disease},}\ }\href@noop {}
  {\bibfield  {journal} {\bibinfo  {journal} {Brain and language}\ }\textbf
  {\bibinfo {volume} {84}},\ \bibinfo {pages} {414--423} (\bibinfo {year}
  {2003})}\BibitemShut {NoStop}%
\bibitem [{\citenamefont {Rogers}(2000)}]{rogers2000interventions}%
  \BibitemOpen
  \bibfield  {author} {\bibinfo {author} {\bibfnamefont {Sally~J}\ \bibnamefont
  {Rogers}},\ }\bibfield  {title} {\enquote {\bibinfo {title} {Interventions
  that facilitate socialization in children with autism},}\ }\href@noop {}
  {\bibfield  {journal} {\bibinfo  {journal} {Journal of autism and
  developmental disorders}\ }\textbf {\bibinfo {volume} {30}},\ \bibinfo
  {pages} {399--409} (\bibinfo {year} {2000})}\BibitemShut {NoStop}%
\bibitem [{\citenamefont {Tennent}\ \emph {et~al.}(2019)\citenamefont
  {Tennent}, \citenamefont {Shen},\ and\ \citenamefont
  {Jung}}]{tennent2019micbot}%
  \BibitemOpen
  \bibfield  {author} {\bibinfo {author} {\bibfnamefont {Hamish}\ \bibnamefont
  {Tennent}}, \bibinfo {author} {\bibfnamefont {Solace}\ \bibnamefont {Shen}},
  \ and\ \bibinfo {author} {\bibfnamefont {Malte}\ \bibnamefont {Jung}},\
  }\bibfield  {title} {\enquote {\bibinfo {title} {Micbot: A peripheral robotic
  object to shape conversational dynamics and team performance},}\ }in\
  \href@noop {} {\emph {\bibinfo {booktitle} {2019 14th ACM/IEEE International
  Conference on Human-Robot Interaction (HRI)}}}\ (\bibinfo {organization}
  {IEEE},\ \bibinfo {year} {2019})\ pp.\ \bibinfo {pages}
  {133--142}\BibitemShut {NoStop}%
\bibitem [{\citenamefont {Nietlisbach}\ and\ \citenamefont
  {Maercker}(2009)}]{nietlisbach2009social}%
  \BibitemOpen
  \bibfield  {author} {\bibinfo {author} {\bibfnamefont {Gabriela}\
  \bibnamefont {Nietlisbach}}\ and\ \bibinfo {author} {\bibfnamefont {Andreas}\
  \bibnamefont {Maercker}},\ }\bibfield  {title} {\enquote {\bibinfo {title}
  {Social cognition and interpersonal impairments in trauma survivors with
  ptsd},}\ }\href@noop {} {\bibfield  {journal} {\bibinfo  {journal} {Journal
  of Aggression, Maltreatment \& Trauma}\ }\textbf {\bibinfo {volume} {18}},\
  \bibinfo {pages} {382--402} (\bibinfo {year} {2009})}\BibitemShut {NoStop}%
\bibitem [{\citenamefont {Weiss}\ \emph {et~al.}(2006)\citenamefont {Weiss},
  \citenamefont {Ramakrishna},\ and\ \citenamefont {Somma}}]{weiss2006health}%
  \BibitemOpen
  \bibfield  {author} {\bibinfo {author} {\bibfnamefont {Mitchell~G}\
  \bibnamefont {Weiss}}, \bibinfo {author} {\bibfnamefont {Jayashree}\
  \bibnamefont {Ramakrishna}}, \ and\ \bibinfo {author} {\bibfnamefont {Daryl}\
  \bibnamefont {Somma}},\ }\bibfield  {title} {\enquote {\bibinfo {title}
  {Health-related stigma: rethinking concepts and interventions},}\ }\href@noop
  {} {\bibfield  {journal} {\bibinfo  {journal} {Psychology, health \&
  medicine}\ }\textbf {\bibinfo {volume} {11}},\ \bibinfo {pages} {277--287}
  (\bibinfo {year} {2006})}\BibitemShut {NoStop}%
\bibitem [{\citenamefont {Dovidio}\ \emph {et~al.}(2000)\citenamefont
  {Dovidio}, \citenamefont {Major},\ and\ \citenamefont
  {Crocker}}]{dovidio2000stigma}%
  \BibitemOpen
  \bibfield  {author} {\bibinfo {author} {\bibfnamefont {John~F}\ \bibnamefont
  {Dovidio}}, \bibinfo {author} {\bibfnamefont {Brenda}\ \bibnamefont {Major}},
  \ and\ \bibinfo {author} {\bibfnamefont {Jennifer}\ \bibnamefont {Crocker}},\
  }\bibfield  {title} {\enquote {\bibinfo {title} {Stigma: Introduction and
  overview.}}\ }\href@noop {} {\  (\bibinfo {year} {2000})}\BibitemShut
  {NoStop}%
\bibitem [{\citenamefont {Van~Brakel}(2006)}]{van2006measuring}%
  \BibitemOpen
  \bibfield  {author} {\bibinfo {author} {\bibfnamefont {Wim~H}\ \bibnamefont
  {Van~Brakel}},\ }\bibfield  {title} {\enquote {\bibinfo {title} {Measuring
  health-related stigma?a literature review},}\ }\href@noop {} {\bibfield
  {journal} {\bibinfo  {journal} {Psychology, health \& medicine}\ }\textbf
  {\bibinfo {volume} {11}},\ \bibinfo {pages} {307--334} (\bibinfo {year}
  {2006})}\BibitemShut {NoStop}%
\bibitem [{\citenamefont {Pettinati}\ \emph {et~al.}(2016)\citenamefont
  {Pettinati}, \citenamefont {Arkin},\ and\ \citenamefont
  {Shim}}]{pettinati2016influence}%
  \BibitemOpen
  \bibfield  {author} {\bibinfo {author} {\bibfnamefont {Michael~J}\
  \bibnamefont {Pettinati}}, \bibinfo {author} {\bibfnamefont {Ronald~C}\
  \bibnamefont {Arkin}}, \ and\ \bibinfo {author} {\bibfnamefont {Jaeeun}\
  \bibnamefont {Shim}},\ }\bibfield  {title} {\enquote {\bibinfo {title} {The
  influence of a peripheral social robot on self-disclosure},}\ }in\ \href@noop
  {} {\emph {\bibinfo {booktitle} {2016 25th IEEE International Symposium on
  Robot and Human Interactive Communication (RO-MAN)}}}\ (\bibinfo
  {organization} {IEEE},\ \bibinfo {year} {2016})\ pp.\ \bibinfo {pages}
  {1063--1070}\BibitemShut {NoStop}%
\bibitem [{\citenamefont {Bethel}\ \emph {et~al.}(2016)\citenamefont {Bethel},
  \citenamefont {Henkel}, \citenamefont {Stives}, \citenamefont {May},
  \citenamefont {Eakin}, \citenamefont {Pilkinton}, \citenamefont {Jones},\
  and\ \citenamefont {Stubbs-Richardson}}]{bethel2016using}%
  \BibitemOpen
  \bibfield  {author} {\bibinfo {author} {\bibfnamefont {Cindy~L}\ \bibnamefont
  {Bethel}}, \bibinfo {author} {\bibfnamefont {Zachary}\ \bibnamefont
  {Henkel}}, \bibinfo {author} {\bibfnamefont {Kristen}\ \bibnamefont
  {Stives}}, \bibinfo {author} {\bibfnamefont {David~C}\ \bibnamefont {May}},
  \bibinfo {author} {\bibfnamefont {Deborah~K}\ \bibnamefont {Eakin}}, \bibinfo
  {author} {\bibfnamefont {Melinda}\ \bibnamefont {Pilkinton}}, \bibinfo
  {author} {\bibfnamefont {Alexis}\ \bibnamefont {Jones}}, \ and\ \bibinfo
  {author} {\bibfnamefont {Megan}\ \bibnamefont {Stubbs-Richardson}},\
  }\bibfield  {title} {\enquote {\bibinfo {title} {Using robots to interview
  children about bullying: Lessons learned from an exploratory study},}\ }in\
  \href@noop {} {\emph {\bibinfo {booktitle} {2016 25th IEEE International
  Symposium on Robot and Human Interactive Communication (RO-MAN)}}}\ (\bibinfo
  {organization} {IEEE},\ \bibinfo {year} {2016})\ pp.\ \bibinfo {pages}
  {712--717}\BibitemShut {NoStop}%
\bibitem [{\citenamefont {Sharkey}\ and\ \citenamefont
  {Sharkey}(2012)}]{sharkey2012granny}%
  \BibitemOpen
  \bibfield  {author} {\bibinfo {author} {\bibfnamefont {Amanda}\ \bibnamefont
  {Sharkey}}\ and\ \bibinfo {author} {\bibfnamefont {Noel}\ \bibnamefont
  {Sharkey}},\ }\bibfield  {title} {\enquote {\bibinfo {title} {Granny and the
  robots: ethical issues in robot care for the elderly},}\ }\href@noop {}
  {\bibfield  {journal} {\bibinfo  {journal} {Ethics and information
  technology}\ }\textbf {\bibinfo {volume} {14}},\ \bibinfo {pages} {27--40}
  (\bibinfo {year} {2012})}\BibitemShut {NoStop}%
\bibitem [{\citenamefont {Wilson}\ \emph
  {et~al.}(2016{\natexlab{c}})\citenamefont {Wilson}, \citenamefont {Lee},
  \citenamefont {Saechao},\ and\ \citenamefont {Scheutz}}]{wilson2016autonomy}%
  \BibitemOpen
  \bibfield  {author} {\bibinfo {author} {\bibfnamefont {Jason~R}\ \bibnamefont
  {Wilson}}, \bibinfo {author} {\bibfnamefont {Nah~Young}\ \bibnamefont {Lee}},
  \bibinfo {author} {\bibfnamefont {Annie}\ \bibnamefont {Saechao}}, \ and\
  \bibinfo {author} {\bibfnamefont {Matthias}\ \bibnamefont {Scheutz}},\
  }\bibfield  {title} {\enquote {\bibinfo {title} {Autonomy and dignity:
  Principles in designing effective social robots to assist in the care of
  older adults},}\ }in\ \href@noop {} {\emph {\bibinfo {booktitle} {Workshop on
  using social robots to improve the quality of life in the elderly, icsr}}}\
  (\bibinfo {year} {2016})\BibitemShut {NoStop}%
\bibitem [{\citenamefont {Arnold}\ and\ \citenamefont
  {Scheutz}(2017{\natexlab{a}})}]{arnold2017beyond}%
  \BibitemOpen
  \bibfield  {author} {\bibinfo {author} {\bibfnamefont {Thomas}\ \bibnamefont
  {Arnold}}\ and\ \bibinfo {author} {\bibfnamefont {Matthias}\ \bibnamefont
  {Scheutz}},\ }\bibfield  {title} {\enquote {\bibinfo {title} {Beyond moral
  dilemmas: Exploring the ethical landscape in hri},}\ }in\ \href@noop {}
  {\emph {\bibinfo {booktitle} {2017 12th ACM/IEEE International Conference on
  Human-Robot Interaction (HRI}}}\ (\bibinfo {organization} {IEEE},\ \bibinfo
  {year} {2017})\ pp.\ \bibinfo {pages} {445--452}\BibitemShut {NoStop}%
\bibitem [{\citenamefont {Wilson}\ \emph
  {et~al.}(2016{\natexlab{d}})\citenamefont {Wilson}, \citenamefont {Scheutz},\
  and\ \citenamefont {Briggs}}]{wilson2016reflections}%
  \BibitemOpen
  \bibfield  {author} {\bibinfo {author} {\bibfnamefont {Jason~R}\ \bibnamefont
  {Wilson}}, \bibinfo {author} {\bibfnamefont {Matthias}\ \bibnamefont
  {Scheutz}}, \ and\ \bibinfo {author} {\bibfnamefont {Gordon}\ \bibnamefont
  {Briggs}},\ }\bibfield  {title} {\enquote {\bibinfo {title} {Reflections on
  the design challenges prompted by affect-aware socially assistive robots},}\
  }in\ \href@noop {} {\emph {\bibinfo {booktitle} {Emotions and Personality in
  Personalized Services}}}\ (\bibinfo  {publisher} {Springer},\ \bibinfo {year}
  {2016})\ pp.\ \bibinfo {pages} {377--395}\BibitemShut {NoStop}%
\bibitem [{\citenamefont {Turkle}\ \emph {et~al.}(2006)\citenamefont {Turkle},
  \citenamefont {Taggart}, \citenamefont {Kidd},\ and\ \citenamefont
  {Dast{\'e}}}]{turkle2006relational}%
  \BibitemOpen
  \bibfield  {author} {\bibinfo {author} {\bibfnamefont {Sherry}\ \bibnamefont
  {Turkle}}, \bibinfo {author} {\bibfnamefont {Will}\ \bibnamefont {Taggart}},
  \bibinfo {author} {\bibfnamefont {Cory~D}\ \bibnamefont {Kidd}}, \ and\
  \bibinfo {author} {\bibfnamefont {Olivia}\ \bibnamefont {Dast{\'e}}},\
  }\bibfield  {title} {\enquote {\bibinfo {title} {Relational artifacts with
  children and elders: the complexities of cybercompanionship},}\ }\href@noop
  {} {\bibfield  {journal} {\bibinfo  {journal} {Connection Science}\ }\textbf
  {\bibinfo {volume} {18}},\ \bibinfo {pages} {347--361} (\bibinfo {year}
  {2006})}\BibitemShut {NoStop}%
\bibitem [{\citenamefont {Arnold}\ and\ \citenamefont
  {Scheutz}(2017{\natexlab{b}})}]{arnold2017tactile}%
  \BibitemOpen
  \bibfield  {author} {\bibinfo {author} {\bibfnamefont {Thomas}\ \bibnamefont
  {Arnold}}\ and\ \bibinfo {author} {\bibfnamefont {Matthias}\ \bibnamefont
  {Scheutz}},\ }\bibfield  {title} {\enquote {\bibinfo {title} {The tactile
  ethics of soft robotics: Designing wisely for human--robot interaction},}\
  }\href@noop {} {\bibfield  {journal} {\bibinfo  {journal} {Soft robotics}\
  }\textbf {\bibinfo {volume} {4}},\ \bibinfo {pages} {81--87} (\bibinfo {year}
  {2017}{\natexlab{b}})}\BibitemShut {NoStop}%
\bibitem [{\citenamefont {Scheutz}(2011)}]{scheutz201113}%
  \BibitemOpen
  \bibfield  {author} {\bibinfo {author} {\bibfnamefont {Matthias}\
  \bibnamefont {Scheutz}},\ }\bibfield  {title} {\enquote {\bibinfo {title} {13
  the inherent dangers of unidirectional emotional bonds between humans and
  social robots},}\ }\href@noop {} {\bibfield  {journal} {\bibinfo  {journal}
  {Robot ethics: The ethical and social implications of robotics}\ ,\ \bibinfo
  {pages} {205}} (\bibinfo {year} {2011})}\BibitemShut {NoStop}%
\bibitem [{\citenamefont {Bartneck}\ \emph {et~al.}(2005)\citenamefont
  {Bartneck}, \citenamefont {Nomura}, \citenamefont {Kanda}, \citenamefont
  {Suzuki},\ and\ \citenamefont {Kato}}]{bartneck2005cultural}%
  \BibitemOpen
  \bibfield  {author} {\bibinfo {author} {\bibfnamefont {Christoph}\
  \bibnamefont {Bartneck}}, \bibinfo {author} {\bibfnamefont {Tatsuya}\
  \bibnamefont {Nomura}}, \bibinfo {author} {\bibfnamefont {Takayuki}\
  \bibnamefont {Kanda}}, \bibinfo {author} {\bibfnamefont {Tomohiro}\
  \bibnamefont {Suzuki}}, \ and\ \bibinfo {author} {\bibfnamefont {Kennsuke}\
  \bibnamefont {Kato}},\ }\bibfield  {title} {\enquote {\bibinfo {title}
  {Cultural differences in attitudes towards robots},}\ }in\ \href@noop {}
  {\emph {\bibinfo {booktitle} {Proc. Symposium on robot companions (SSAISB
  2005 convention)}}}\ (\bibinfo {year} {2005})\ pp.\ \bibinfo {pages}
  {1--4}\BibitemShut {NoStop}%
\end{thebibliography}%

\end{document}